\documentclass[review,12pt,double space]{article}
\usepackage[utf8]{inputenc}
\usepackage{amsmath}
\usepackage{hyperref}
\usepackage[nameinlink,capitalize]{cleveref}
\usepackage{graphicx}
\usepackage{xcolor}
\usepackage{soul}

\newcommand{\etal}{\textit{et al. }}

\title{Lithium-ion battery degradation: how to model it}
\author{Simon E. J. O'Kane\textsuperscript{1,6,a}, Weilong Ai\textsuperscript{2,6,b}, Ganesh Madabattula\textsuperscript{1,6,c},\\Diego Alonso Alvarez\textsuperscript{3,6}, Robert Timms\textsuperscript{4,6}, Valentin Sulzer\textsuperscript{5,6},\\Jacqueline Sophie Edge\textsuperscript{1,6}, Billy Wu\textsuperscript{2,6}, Gregory J. Offer\textsuperscript{1,6}, Monica Marinescu\textsuperscript{1,6}\\\\\textsuperscript{1}Department of Mechanical Engineering, Imperial College London, UK\\\textsuperscript{2}Dyson School of Design Engineering, Imperial College London, UK\\\textsuperscript{3}Research Computing Service, ICT, Imperial College London, UK\\\textsuperscript{4}Mathematical Institute, University of Oxford, UK\\\textsuperscript{5}Department of Mechanical Engineering, Carnegie Mellon University, USA\\\textsuperscript{6}The Faraday Institution, UK\\\\\textsuperscript{a}Email: s.okane@imperial.ac.uk\\\textsuperscript{b}Current address: School of Civil Engineering, Southeast University, PR China\\\textsuperscript{c}Current address: BritishVolt Ltd., UK}
\date{25th January 2022}

\begin{document}

\maketitle

\begin{abstract}
    Predicting lithium-ion battery degradation is worth billions to the global automotive, aviation and energy storage industries, to improve performance and safety and reduce warranty liabilities. However, very few published models of battery degradation explicitly consider the interactions between more than two degradation mechanisms, and none do so within a single electrode. In this paper, the first published attempt to directly couple more than two degradation mechanisms in the negative electrode is reported. The results are used to map different pathways through the complicated path dependent and non-linear degradation space. Four degradation mechanisms are coupled in PyBaMM, an open source modelling environment uniquely developed to allow new physics to be implemented and explored quickly and easily. Crucially it is possible to see `inside’ the model and observe the consequences of the different patterns of degradation, such as loss of lithium inventory and loss of active material. For the same cell, five different pathways that can result in end-of-life have already been found, depending on how the cell is used. Such information would enable a product designer to either extend life or predict life based upon the usage pattern. However, parameterization of the degradation models remains as a major challenge, and requires the attention of the international battery community.
\end{abstract}

\section*{Broader context}
Lithium-ion batteries were originally developed for portable electronics, for which energy and power density are of greater concern than lifetime. With their expansion into transport and energy storage applications, lifetime has become a significant concern. Many different degradation mechanisms occur inside LIBs and none of them can be measured directly during operation, only their consequences can be ‘observed’ as capacity and power fade. In contrast, computer simulations allow access to the internal states of the battery and track the evolution of each relevant degradation mechanism. While many models of LIB degradation exist, only few consider more than two mechanisms, and even fewer consider direct interactions between the mechanisms. Yet, degradation in a real battery involves multiple mechanisms that are strongly coupled with each other. In this work, the open-source modelling environment Python Battery Mathematical Modelling (PyBaMM) is modified to include four different degradation mechanisms, all of which directly interact with each other. Being open-source makes it possible for other researchers to easily reproduce and build upon this work, adding more degradation mechanisms and studying their interactions. This is therefore an important step on the path towards a universal model for lithium-ion battery degradation.

\section{Introduction}\label{intro}
Lithium-ion batteries (LiBs) have already transformed our world by triggering a revolution in portable electronics. They are now enabling further transformations in electric vehicles (EVs) and stationary energy storage applications \cite{Gur2018}. However, in these applications, the batteries are operated in harsher conditions and required to last longer. As a result, cycle life and safety have become the main concern, prompting a significant increase in research into the physics of battery degradation. A comprehensive review of LiB degradation was published in 2005 \cite{Vetter2005}, but the expansion in LiB applications has prompted many more reviews in recent years, each with a different focus. Hendricks \etal \cite{Hendricks2015} and Birkl \etal \cite{Birkl2017} focus on diagnosing degradation, Han \etal \cite{Han2019} and Tomaszewska \etal \cite{Tomaszewska2019} focus on the effect of cycling conditions, and Edge \etal \cite{Edge2021} focus on the interactions between different degradation mechanisms.\\\\
While experiments, both destructive and life cycle testing, provide essential information, much can be achieved by models, as long as they are proved to be trustworthy in their prediction of degradation. It is important to include the interactions between degradation mechanisms, as behaviour of combined mechanisms cannot always be reproduced by combining individual mechanisms. Empirical models may capture this combined behaviour, but they can offer insights into how to prevent degradation only when combined with physics-based models.\\\\
In their review of interactions between degradation mechanisms, Edge \etal \cite{Edge2021} reported that several computational studies have attempted to analyse some of these interactions, but few included more than two mechanisms. Lin \etal \cite{Lin2013a} modelled solid-electrolyte interphase (SEI) growth on the negative electrode, Mn dissolution from the positive electrode, Mn deposition on the negative electrode and hydrogen gas evolution. Jin \etal \cite{Jin2018ApplicabilityDesign} modelled SEI growth and mechanical loss of active material in both electrodes. Reniers, Mulder and Howey \cite{Reniers2019ReviewBatteries} modelled SEI growth, lithium plating and mechanical loss of active material in the negative electrode, and Mn dissolution from the positive electrode. Li, Landers and Park modelled SEI growth on cracks in the negative electrode particles, alongside the Mn dissolution equations of Lin \etal. Keil and Jossen \cite{Keil2020} focused on degradation at the negative electrode, modelling partially reversible lithium plating and SEI growth on cracks.\\\\
Crucially, none of these studies consider the direct interactions between more than two mechanisms in the same electrode. Instead, they focused on indirect interactions, where each mechanism affects the underlying battery model, which, in turn, affects the other mechanisms. Sections \ref{indirect} and \ref{direct} explore this distinction in-depth.\\\\
In this work, a Doyle-Fuller-Newman (DFN) model is created that includes strongly coupled models of degradation in the graphite negative electrode and is implemented in PyBaMM \cite{Sulzer2020PythonPyBaMM}. The interaction between Li plating and the SEI is modelled by letting plated Li decay into inactive ``dead lithium" over time. The thicker the SEI, the slower the rate of dead lithium formation. Electrode crack propagation and particle fracture are unified into a single stress-based model \cite{Ai2020ElectrochemicalCells}. Solvent diffusion-limited SEI growth occurs on the crack surfaces as they propagate \cite{Single2018IdentifyingInterphase}.

\section{Degradation mechanisms}\label{mechanisms}
The degradation mechanisms specific to the graphite negative electrode are generally better understood than those specific to the positive electrode, making the negative electrode the main focus of this work. As particle fracture is known to occur in both electrodes, this one degradation mechanism is also simulated at the positive electrode side.
\subsection{SEI layer growth}
The organic electrolytes used in most LiBs are unstable at voltages below 1 V vs. Li/Li\textsuperscript{+} \cite{Wang2016}. Lithiated graphite has an equilibrium potential well below this threshold, causing the electrolyte to break down, reacting with the lithiated graphite and forming an SEI layer on the graphite surface. \\\\
The SEI contributes to battery degradation in several ways. The compounds in the SEI contain lithium, trapping it and preventing it from participating in charge storage; an effect known as loss of lithium inventory. The SEI also increases impedance in a more direct way; while Li\textsuperscript{+} ions can move through the SEI, its ionic resistance is higher than that of the electrolyte.\\\\
The model developed by Safari \etal \cite{Safari2009MultimodalBatteries} has become the basis for most SEI models in the current literature. Safari \etal noted that the main SEI reaction, in which Li\textsuperscript{+} reacts with one of the organic solvents, is limited by two factors. First, the solvent molecule must diffuse through the existing SEI to reach the graphite surface. Then, the graphite must supply an electron to reduce the solvent molecule. Safari \etal's model accounts for both these limitations, but found that the diffusion-limited regime provides the best fit to experimental data. A simplified diffusion-limited model is therefore used in this work.
\subsection{Lithium plating}\label{plating}
Another side reaction that occurs on the negative electrode is lithium plating, where Li\textsuperscript{+} ions from the electrolyte form Li metal on the graphite surface, instead of intercalating into the graphite bulk. Li plating has been reviewed in dedicated papers \cite{Liu2016,Waldmann2018} and as part of a wider review of fast charging \cite{Tomaszewska2019}. It occurs under conditions that slow down the main intercalation reaction, such as low temperatures or when the graphite surface is fully lithiated, making the competing Li plating reaction more favourable. Conditions that raise the electrolyte potential to unusually high levels, such as fast charging or pore clogging, also cause Li plating.\\\\
As with other electroplating reactions, there exists a reverse process, termed Li stripping, in which the Li metal atoms are oxidized and return to the electrolyte as Li\textsuperscript{+} ions. However, the surface of the Li metal exposed to the electrolyte reacts with it to form SEI, so not all of the lithium can be recovered via stripping. SEI growth atop plated Li can cause parts of the Li metal deposit to become electrically isolated, forming "dead lithium" that cannot be recovered \cite{Ren2018, Zhao2019Electrochemical-thermalTemperatures}.\\\\
The first model of Li plating and stripping on graphite was reported by Arora, Doyle and White \cite{arora_mathematical_1999-3}, who used a Butler-Volmer equation for plating/stripping. However, there is no dependence on the amount of plated Li in their model, causing it to predict negative values of the plated Li when all Li is stripped. This means the model can only be used for charge and not for discharge. Despite this limitation, the model received experimental validation from Ge \etal \cite{Ge2017}.\\\\
The three-way interaction between Li plating, stripping and SEI formation has resulted in a range of models in recent years. Yang \etal \cite{Yang2017ModelingAging} used a simplified version of Safari \etal's \cite{Safari2009MultimodalBatteries} model of SEI growth and a simple Tafel equation for irreversible Li plating. Their follow-up paper \cite{Yang2018ACells} introduced an updated Butler-Volmer equation including dependence on the amount of plated Li, taken from work on lithium metal batteries \cite{Chen2017}. Ren \etal \cite{Ren2018} and von L\"{u}ders \etal \cite{vonLuders2019} used a different approach, multiplying Arora, Doyle and White's original equation by a function that becomes zero when all the lithium is stripped, preventing the plated Li from going negative. Zhao \etal \cite{Zhao2019Electrochemical-thermalTemperatures} let a preset fraction of plated Li turn into irreversible SEI instantaneously upon plating. Keil and Jossen \cite{Keil2020} made this fraction time-dependent to achieve an excellent fit to experimental data. However, no models exist where plated Li gradually turns into SEI over time. A first step towards such a model is proposed in this work.
\subsection{Particle fracture}\label{fracture}
The volume of the electrode changes substantially during cycling, e.g. for graphite it expands on lithiation and contracts on delithiation. Concentration gradients inside the particles therefore cause mechanical stress, which can cause cracks to develop \cite{Deshpande2012a}. Cracking occurs more severely during higher currents and for larger particle sizes \cite{Christensen2006}. Towards the end of battery life, complete fracture can occur, often close to the separator, due to the higher local reaction current density during cycling \cite{Xu2019}.\\\\
Electrode particle fracture causes degradation in three ways. The cracks reveal additional surface area for SEI growth, causing further loss of lithium inventory \cite{Li2018}. The conductive additives can also detach from the particles during delithiation \cite{Xu2019}, leading to a loss in electronic conductivity. Complete fracture or complete detachment from the binder causes the particle to become electrically isolated, causing loss of active material \cite{Delacourt2012,Laresgoiti2015}.\\\\
Different models for particle cracking have been proposed, depending on the length scale of the problem. At cell level, a fatigue crack model (Paris' model) has been introduced to the single particle model (SPM) and coupled with the SEI formation and growth \cite{Li2018}. At electrode level, the morphology of LiNi$_x$Mn$_y$Co$_z$O$_2$ positive electrodes has been studied by Xu \etal \cite{Xu2019}, using finite element methods in 3D. At particle level, Laresgoiti \etal \cite{Laresgoiti2015} calculated the mechanical stress inside the graphite and SEI separately. The model presented in this work combines these three relevant models, while keeping the computational load manageable, as explained in Sections \ref{crack model} and \ref{LAM model}
\subsection{Other degradation mechanisms}
Degradation mechanisms specific to the positive electrode are less well understood, but are known to include irreversible phase changes \cite{Jung2017}, transition metal dissolution and formation of a positive SEI (pSEI) layer on the cathode particles. Transition metals, especially Mn, can also migrate to the negative electrode and be deposited there in a similar process to lithium plating \cite{Blyr1998}. Unlike SEI growth or Li plating, new models are required, such as that developed for irreversible phase changes in NMC positive electrodes \cite{Ghosh2021}. Including these mechanisms remains the topic of future work.\\\\
Another factor neglected in this work is the increasing use of composite electrodes with more than one type of particle. In electrodes with a bimodal size distribution, the particles can be classified as small and large \cite{Ecker2015}, which impacts mechanical degradation. The increasing use of graphite negative electrodes with a small fraction of SiO\textsubscript{x} particles also presents a challenge to modelling mechanical degradation. These electrodes have higher capacity but show strong hysteresis \cite{Jiang2020VoltageAmorphization} and are highly vulnerable to mechanical degradation due to the large volume change of the SiO\textsubscript{x} particles on lithiation.

\subsection{Indirect interactions}\label{indirect}
All mechanisms contribute to five degradation modes \cite{Edge2021}: loss of lithium inventory (LLI), loss of active material (LAM) in the negative and positive electrodes, stoichiometric drift and impedance change. SEI formation causes LLI by immobilising Li\textsuperscript{+} ions and impedance changes via film resistance and pore clogging \cite{Safari2009MultimodalBatteries}. Lithium plating causes LLI in the most literal sense by forming dead lithium and also causes some impedance changes via pore clogging \cite{Liu2016}. Particle fracture can contribute to all five modes: LLI by enabling additional SEI formation on cracks \cite{Li2020a}, impedance changes including binder degradation \cite{Xu2019} and LAM in the extreme case when particles detach completely \cite{Delacourt2012,Laresgoiti2015}.\\\\
The five degradation modes affect the degradation mechanisms, resulting in feedback loops, both positive and negative \cite{Edge2021}. LLI results in negative feedback. Having less cyclable lithium inhibits lithium plating, by preventing the negative electrode from reaching the highly lithiated state in which plating occurs \cite{Reniers2019ReviewBatteries, Gao2021}. In contrast, impedance increase caused by pore clogging is known to aggravate Li plating by reducing the electrolyte potential \cite{Liu2020}. LAM also triggers a positive feedback loop, as it enhances particle fracture: LAM decreases the interface area between the active material and electrolyte, thus increasing the interfacial current density and, in turn, leading to larger concentration gradients and increased mechanical stress \cite{Reniers2019ReviewBatteries}.
\subsection{Direct interactions}\label{direct}
It is generally acknowledged that SEI growth is self-limiting due to the diffusion limitation of solvent molecules through the SEI. The interactions between SEI growth and other degradation mechanisms, however, are less well-researched. As mentioned in Section \ref{plating}, the three-way interaction between Li plating, Li stripping and SEI growth has resulted in a range of models, none of which form a complete physical description. Another major interaction is the growth of SEI on fresh surface exposed by particle cracking, which has been shown to be faster than SEI growth on pre-existing surface with pre-existing SEI. \cite{Li2020a}\\\\
Most relevant models treat crack-assisted SEI growth and mechanically-induced LAM as separate mechanisms, governed by empirical equations \cite{Jin2018ApplicabilityDesign,Delacourt2012,Ekstrom2015Cell}. In reality they cannot be independent because they have a common origin: the mechanical stress caused by concentration gradients \cite{Laresgoiti2015}. Crack growth and mechanical LAM are different outcomes of the same fundamental physics, so ideally should be modelled in a coupled way.\\\\
A strongly coupled model of battery degradation is presented that seeks to unify the two different forms of mechanical degradation into a single stress model, while also including the direct interactions between SEI growth, Li plating and particle cracking.
\section{Governing equations}\label{equations}
The Doyle-Fuller-Newman (DFN) model of LiBs is chosen for representing the beginning of life behaviour of the battery \cite{Fuller1994SimulationCell}. The system of coupled differential and algebraic equations is listed in the Supplementary Information.
\subsection{SEI growth model}\label{SEI model}
The solvent diffusion limited SEI growth model \cite{Single2018IdentifyingInterphase} is used in this work. In this model, the diffusion of solvent molecules through the SEI layer limits its growth. It assumes steady state, such that the flux of the solvent molecules follows Fick's law,
\begin{gather}
    N_\text{sol}=-D_\text{sol}(T)\frac{\partial c_\text{sol}}{\partial l},\\
    c_\text{sol} =0 \quad \text{at} \quad l=0 ,\\
    c_\text{sol}=c_\text{sol,0} \quad \text{at} \quad l=L_\text{SEI},
\end{gather}
where $c_\text{sol}$ is the solvent concentration, $c_\text{sol,0}$ is the 
solvent concentration in the electrolyte, $D_\text{sol}(T)$ is the solvent diffusion coefficient, $l$ is one location of the SEI layer and $L_\text{SEI}$ is the thickness (and also the location of the outer surface) of the SEI layer. The solution is
\begin{equation}
    c_\text{sol}=\frac{lc_\text{sol,0}}{L_\text{SEI}} \quad \text{and} \quad {N_\text{sol}}=-\frac{c_\text{sol,0} D_\text{sol}(T)}{L_\text{SEI}}.
\end{equation}
By conservation of mass, the interfacial flux density in the SEI layer is
\begin{equation}
    N_\text{SEI} = -N_\text{sol}=\frac{c_\text{sol,0} D_\text{sol}(T)}{L_\text{SEI}}.
\end{equation}
The thickness growth rate of the SEI layer is 
\begin{equation}\label{SEI growth}
    \frac{\partial L_\text{SEI}}{\partial t}=-\tfrac{1}{2}N_\text{sol} \bar{V}_\text{SEI}=\frac{c_\text{sol,0}D_\text{sol}(T)\bar{V}_\text{SEI}}{2L_\text{SEI}}.
\end{equation}
Equation \eqref{SEI growth} shows that the thickness of the SEI layer is determined by the square root of time. The key parameter is $D_\text{sol}(T)$, which is assumed to have an Arrhenius temperature dependence:
\begin{equation}
    D_\text{sol}(T)=D_\text{sol}(T_\text{ref})\exp\left(-\frac{E_\text{sol}}{RT}+\frac{E_\text{sol}}{RT_\text{ref}}\right),
\end{equation}
where $T_\text{ref}=298.15$ K (25 $^\circ$C) throughout. The SEI layer is assumed to have an Ohmic resistivity $\rho_\text{SEI}$, causing a voltage drop $\eta_\text{SEI}$:
\begin{equation}\label{eta_SEI}
    \eta_\text{SEI}=\rho_\text{SEI}L_\text{SEI}\frac{j_\text{tot}}{a_-},
\end{equation}
where $j_\text{tot}$ is the interfacial current density and is defined in Table S1. This SEI growth model is a simplified version of the one used by PyBaMM, which considers two SEI layers as opposed to the single layer considered here. Details of the double-layer model are given in the Supplementary Information.
\subsection{Li plating model}\label{plating model}
The first DFN model of Li plating/stripping on graphite electrodes was published by Arora, Doyle and White \cite{arora_mathematical_1999-3}, whose Butler-Volmer equation had no dependence on the concentration $c_\mathrm{Li}$ of plated Li. Yang \etal \cite{Yang2018ACells} used an updated equation for the Li stripping flux $N_\mathrm{Li}$ that included concentration dependence:
\begin{equation}\label{BV plating 1}
N_\mathrm{Li}=k_\mathrm{Li}(c_\mathrm{Li}^*)^{\alpha_{c,Li}}(c_\mathrm{e}^*)^{\alpha_{a,Li}}\left(\frac{c_\mathrm{Li}}{c_\mathrm{Li}^*}\exp\left(\frac{\alpha_\mathrm{a,Li}F}{RT}(\phi_\mathrm{s}-\phi_\mathrm{e})\right)-\frac{c_\mathrm{e}}{c_\mathrm{e}^*}\exp\left(-\frac{\alpha_\mathrm{c,Li}F}{RT}(\phi_\mathrm{s}-\phi_\mathrm{e})\right)\right),
\end{equation}
where $c_\mathrm{e}^*=c_\mathrm{eq}$ and both $k_\mathrm{Li}$ and $c_\mathrm{Li}^*$ were used as fitting parameters. The transfer coefficients $\alpha_\mathrm{a,Li}$ and $\alpha_\mathrm{c,Li}$ were set to 0.3 and 0.7 respectively, as Arora, Doyle and White \cite{arora_mathematical_1999-3} did. O'Kane \etal \cite{OKane2020} used a simpler expression taken from Wood \etal \cite{Wood2016DendritesMicroscopy}:
\begin{equation}\label{BV plating 2}
N_\mathrm{Li}=k_\mathrm{Li}\left(c_\mathrm{Li}\exp\left(\frac{F(\phi_\mathrm{s}-\phi_\mathrm{e})}{2RT}\right)-c_\mathrm{e}\exp\left(-\frac{F(\phi_\mathrm{s}-\phi_\mathrm{e})}{2RT}\right)\right).
\end{equation}
Whichever form of the Butler-Volmer equation is used, $c_\mathrm{Li}(x,t)$ is found by solving the differential equation
\begin{equation}\label{c_Li}
\frac{\partial c_\mathrm{Li}}{\partial t}=-a_-N_\mathrm{Li}.
\end{equation}
However, O'Kane \etal's model is not a complete description of Li plating, as it does not consider the subsequent decay of plated Li into SEI and "dead lithium". As previously mentioned, the three-way interaction between plating, stripping and SEI formation results in a range of models throughout the literature. None of these models are strongly coupled and so a new model, where plated Li decays into dead Li over time, is proposed here. This time-limited decay adds a second term to \eqref{c_Li}:
\begin{equation}\label{c_Li with decay}
\frac{\partial c_\mathrm{Li}}{\partial t}=-a_-N_\mathrm{Li}-\gamma c_{Li}.
\end{equation}
The concentration $c_\mathrm{dl}$ of dead lithium is given by
\begin{equation}\label{c_dl}
\frac{\partial c_\mathrm{dl}}{\partial t}=\gamma c_{Li},
\end{equation}
where the decay rate $\gamma$ has dimensions of s\textsuperscript{-1}. Assuming that a reaction with the electrolyte solvent is required to turn plated Li into dead Li, $\gamma$ is not constant; instead, it is diffusion-limited just like the SEI formation reaction, making it dependent on the SEI thickness $L_\mathrm{SEI}$. A simple way to model this dependence is 
\begin{equation}\label{gamma}
\gamma(L_\mathrm{SEI})=\gamma_0\frac{L_\mathrm{SEI,0}}{L_\mathrm{SEI}},
\end{equation}
where $L_\mathrm{SEI,0}$ is the SEI thickness at $t=0$ and $\gamma_0$ is a fitting parameter. Returning to the Butler-Volmer equation for $N_\mathrm{sr}$, \eqref{BV plating 2} is updated with variable $\alpha_\mathrm{a,Li}$ and $\alpha_\mathrm{c,Li}$ and the SEI overpotential $\eta_{SEI}$, resulting in
\begin{equation}\label{BV plating 3}
N_\mathrm{Li}=k_\mathrm{Li}\left(c_\mathrm{Li}\exp\left(\frac{F\alpha_\mathrm{a,Li}\eta_\mathrm{Li}}{RT}\right)-c_\mathrm{e}\exp\left(-\frac{F\alpha_\mathrm{c,Li}\eta_\mathrm{Li}}{RT}\right)\right),
\end{equation}
where
\begin{equation}\label{eta_Li}
\eta_{Li}=\phi_s-\phi_e-\eta_\mathrm{SEI}.
\end{equation}
Equations \eqref{c_Li with decay}-\eqref{eta_Li} define the complete Li plating model used in this work.
\subsection{Particle cracking model}\label{crack model}
Electrode materials experience large volume changes during (de)lithiation, and the resulting stress can cause particle cracking, which creates new surfaces and accelerates side reactions including the SEI growth and Li plating. Several crack models have been developed for battery degradation, which are classified as empirical models \cite{Ekstrom2015Cell} and physics-based models \cite{Purewal2014a,Deshpande2012a}. While empirical models are only accurate up to moderate C-rates, e.g. 1 C in Ekstrom and Lindbergh \cite{Ekstrom2015Cell}, physical models can be more accurate for higher C-rates. The crack model in this work is based on the fatigue crack model in Deshpande et al. \cite{Deshpande2012a} and demonstrated below. 

A stress model at particle level was proposed by Zhang et al. \cite{Zhang2007NumericalParticles}, based on the equilibrium of stresses for a free standing spherical electrode particle. The analytical solutions for the radial stress $\sigma_\text{r}$, tangential stress $\sigma_\text{t}$ and displacement $u$ are shown below, respectively:
\begin{subequations}
	\begin{gather}
	\sigma_\text{r}=\frac{2\Omega E}{(1-\nu)}[c_\text{avg}(R_i)-c_\text{avg}(r)],\label{sigma_r}\\
	\sigma_\text{t}=\frac{\Omega E}{(1-\nu)}[2c_\text{avg}(R_i)+c_\text{avg}(r)-\bar{c}/3),\label{sigma_t}\\
	u=\frac{(1+\nu)}{(1-\nu)}\Omega rc_\text{avg}(r)+\frac{2(1-2\nu)}{(1-\nu)}\Omega rc_\text{avg}(R_i),\label{u}
	\end{gather}
\end{subequations}   
where $\Omega$ is the partial molar volume, $E$ is the Young's modulus, $\nu$ is the Possion's ratio, $R_i$ is the radius of the particle and $c_\text{avg}(r)$ is the average Li\textsuperscript{+} concentration between 0 and $r$:
\begin{equation}
    c_\text{avg}(r)=\frac{1}{3r^3}\int^r_0 \bar{c}r^2\,\text{d}r,
\end{equation}
where $\bar{c}=c-c_\text{ref}$ is the departure in lithium concentration from the reference value $c_\text{ref}$ for the stress-free case. The magnitude of stress is determined by the lithium concentration gradient and particle radius, as shown in \eqref{sigma_r}-\eqref{u}. Both $\sigma_\text{r}$ and $\sigma_\text{t}$ are defined as being positive for tensile stress and negative for compressive stress. This stress model has been incorporated into the P2D model for battery pouch cells \cite{Ai2020ElectrochemicalCells} and was used to predict the thickness evolution of the pouch cell.

Electrode particles experience a cyclic stress loading during multiple charges and discharges, which can lead to fatigue cracking. However, it is very challenging to track the crack patterns experimentally. The assumptions for cracks in Deshpande \etal \cite{Deshpande2012a} are applied here: identical micro cracks on the electrode particle surface with a length $l_\text{cr}$, a width $w_\text{cr}$ and a density of the crack number per unit area $\rho_\text{cr}$; cracks grow in length during cycling but their width and density is constant. The fatigue crack growth model follows Paris' law \cite{Deshpande2012a}   
\begin{equation}
\frac{\text{d}l_\text{cr}}{\text{d}N}=\frac{k_\text{cr}}{t_0}(\sigma_\text{t}b_\text{cr}\sqrt{\pi l_\text{cr}})^{m_\text{cr}}\quad\sigma_\text{t}>0,
\end{equation}
where $t_0$ is the time for one cycle, $b_\text{cr}$ is the stress intensity factor correction, $k_\text{cr}$ and $m_\text{cr}$ are constants that are determined from experimental data, e.g. the approach in Purewal et al. \cite{Purewal2014a}. The condition $\sigma_\text{t}>0$ means only tensile stress contributes to crack growth. The instantaneous rate of change of the crack area to volume ratio can therefore be estimated by
\begin{equation}\label{eq:crackingrate}
\frac{\text{d}a_\text{cr}}{\text{d}t} = \frac{a_\pm \rho_\text{cr} w_\text{cr}}{t_0}\cdot\frac{\text{d}l_\text{cr}}{\text{d}t} = \frac{a_\pm \rho_\text{cr} w_\text{cr}}{t_0}\cdot k_\text{cr}(\sigma_\text{t}b_\text{cr}\sqrt{\pi l_\text{cr}})^{m_\text{cr}}\quad\sigma_\text{t}>0.
\end{equation}
For interactions between the SEI growth and particle cracking, one solution is to apply the same SEI growth model on the cracks. However, the SEI formation on the increasing fresh crack surfaces must be considered, which is different from the SEI growth on normal particle surfaces. A very thin initial SEI layer is defined on the cracks, e.g. $L_\text{SEI,cr0} = L_\text{SEI,0}/10000$, leading to fast initial SEI layer growth to simulate the SEI formation stage. The SEI layer thickness is not uniform along cracks, because crack propagation leads to different exposure times for different interface locations along a crack. The averaged thickness of the SEI layer on cracks $L_\text{SEI,cr}$ is used for simplicity, with a time evolution defined by
\begin{equation}\label{L_SEI_cr}
	    \frac{\partial L_\text{SEI,cr}}{\partial t}=\frac{c_\text{sol,0}D_\text{sol}(T)\bar{V}_\text{SEI}}{2L_\text{SEI,cr}} + \frac{\partial l_\text{cr}}{\partial t}\frac{L_\text{SEI,cr0} - L_\text{SEI,cr}}{l_\text{cr}}.
\end{equation}
The first term on the right hand side of \eqref{L_SEI_cr} is always positive and accounts for the diffusion-limited growth of the SEI on existing cracks. The second term is always negative and accounts for the particle crack propagation, which increases the total surface area and therefore lowers the average thickness of the SEI layer on cracks.
\subsection{Loss of active material model}\label{LAM model}
Li-ion batteries suffer from LAM as a result of cycling, either due to electrochemical reactions between the electrodes and the electrolyte, e.g. positive electrode dissolution \cite{Kindermann2017}, or due to mechanical damage from stresses in the electrode material leading to particle cracking and binder detachment. In the current model, only LAM as a consequence of particle cracking is included, as electrode dissolution is not modelled.

The key equations for LAM due to particle cracking are given below, whilst for the detailed derivations the reader is referred to Laresgoiti \etal \cite{Laresgoiti2015} and Reniers, Mulder and Howey \cite{Reniers2019ReviewBatteries}. Using the stress model in \eqref{sigma_r}-\eqref{u}, the decrease in the accessible volume fraction of active materials $\epsilon_\text{a}$ is estimated by
\begin{equation}\label{eq:LAM_N}
    \frac{\partial\epsilon_\text{a}}{\partial t}=\frac{\beta}{t_0}\left( \frac{\sigma_\text{h,max}-\sigma_\text{h,min}}{\sigma_\text{c}}\right)^{m_2}\quad\sigma_\text{h,min}>0,
\end{equation}
where $\beta$ and $m_2$ are two constants normally obtained from experiments, the hydrostatic stress $\sigma_h=(\sigma_r+2\sigma_t)/3$ is a combination of \eqref{sigma_r} and \eqref{sigma_t}, $\sigma_c$ is the critical stress and the subscripts min and max are the minimum and the maximum values respectively. As with particle cracking, only tensile stress ($\sigma_\text{h}>0$) contributes to LAM. Between complete cycles of charge and discharge, the particle can reach a steady state with no stress, i.e. $\sigma_\text{h,min}=0$ and \eqref{eq:LAM_N} can be modified for instantaneous reactions to
\begin{equation}
   \label{eq:LAM_t}
    \frac{\partial\epsilon_\text{a}}{\partial t}=\frac{\beta}{t_0}\left( \frac{\sigma_\text{h}}{\sigma_\text{c}}\right)^{m_2}\quad\sigma_\text{h}>0.
\end{equation}
\section{Simulation methods}
\subsection{Parameters}
The LG M50T cylindrical cell was chosen for simulations because a parameter set suitable for the DFN model is available in Chen \etal \cite{Chen2020a}. The negative electrode consists of graphite with 10$\%$ SiO\textsubscript{x} by mass; the positive electrode consists of NMC 811. The used parameter values are listed in the Supplementary Information.\\\\
PyBaMM contains several different thermal models, any of which can be added to the electrochemical models to create thermally coupled battery models. However, the authors are not aware of sufficient experimental data to parametrize these thermal models for the LG M50 cell. Sturm \etal \cite{Sturm2019} obtained data for the entropic (reversible) heating coefficients as functions of stoichiometry, but did not measure any thermal conductivities or heat capacities. Due to this lack of data, no thermal model is included in this work and the temperature is therefore assumed to be constant.
\subsection{Standard cycling protocol}
Unless otherwise stated, the following cycling protocol is used. First, the cell is held at 4.2~V until the current falls below 50~mA (C/100), followed by a 4~hour rest period. Next the first characterization cycle is applied, consisting of a 0.5~A (C/10) discharge to 2.5~V, then a 1.5~A (0.3C) constant current charge to 4.2~V and finally a constant voltage charge at 4.2~V until the current falls below 50~mA (C/100). The cell is then cycled 1000 times, starting with a 5~A (1C) discharge until 2.5~V and followed by the same 1.5~A/4.2~V constant current/constant voltage charge as in the characterization cycle. After 1000 cycles are complete, a second characterization cycle is applied, identical to the first one. The cell temperature is set to 25 $^\circ$C during the 1000 degradation cycles unless otherwise stated, and is always set to 25 $^\circ$C during the characterization cycles. The procedure described in this paragraph is from here on referred to as ``the standard cycling protocol".\\\\
As the capacity of the simulated cell decreases with ageing, the amount of charge passed during a cycle also decreases. Measures of degradation, such as capacity loss, are plotted against the total charge throughput $Q_\text{tot}$, where
\begin{equation}
    Q_\text{tot}(t) = \int_0^t |I(t)|\,\mathrm{d}t^\prime,
\end{equation}
with $|I(t)|$ the absolute value of the applied current. Simulations with various combinations of fully coupled degradation models are run using the standard cycling protocol and discussed in Section \ref{RaD}.
\subsection{PyBaMM}\label{pybamm}
PyBaMM (Python Battery Mathematical Modelling) is a multiphysics battery modelling software package designed to consolidate the myriad of models in the field and enable the international community of battery modellers to collaborate more effectively. \cite{Sulzer2020PythonPyBaMM} The software has been designed to be extendable and modular so that new models and numerical methods can be easily added and rigorously tested. During the creation of the model proposed here, submodels for the degradation mechanisms of lithium plating, loss of active material and particle cracking were developed and added to the code, alongside SEI growth which was already in PyBaMM. Crucially, coupling between these degradation mechanisms was also implemented.\\\\
The model described in Section \ref{equations} is solved by PyBaMM using the method of lines. First, the equations are discretized in space, using a finite volume discretization with X grid points in the electrodes and separator and Y grid points in the particles. This yields a system of differential-algebraic equations, which is then solved using the IDA solver from SUNDIALS via CasADi \cite{hindmarsh2005sundials, Andersson2018}.\\\\
The model for the constant current discharge and charge phases is the standard DFN model described in the Supporting Information, while for the constant voltage phase an additional algebraic equation for the current is introduced so that the voltage constraint is satisfied. Rest phases are modeled as constant current with the current set to zero. The final state of each phase is used as the initial state of the following phase; when switching between constant current and constant voltage, this requires carefully constructing the new vector of (spatially-discretized) initial conditions, since the set of variables is different for the two models. An alternative, and potentially faster, implementation would be to use a differential equation for the current for the charging phase that will ensure CCCV charging \cite{mohtat2021algorithmic}, but this was not implemented in this work.\\\\
To determine when to switch between phases, PyBaMM keeps track of the `events' (such as a voltage cut-off) as the simulation progresses. While some DAE solvers come with in-built functionality to find events, the CasADi solver used in this work does not. Therefore, PyBaMM uses a `step-and-check' approach to find events, where in each iteration the model is solved for a fixed time window and check whether any events have been crossed. The size of time window requires a trade-off: if the time window is too small, the solver will be slow as it will require many iterations, but if the time window is too large, the solver might step too far past an event and fail.\\\\
If the solver fails during the time window, the window is halved before trying again. If the model is solved successfully and no events have been crossed, the solution is accepted and the time window moved forward.\\\\
If any events have been crossed, the times just before and after the first event crossing are recorded. The simulation is then run between those times with very small steps in order to accurately identify the time at which the event was crossed, which is crucial for keeping track of the discharged capacity as the battery degrades. The additional fine-grain solve is required, instead of interpolation, since the voltage is typically very non-linear close to cut-offs.

\section{Results and Discussion}\label{RaD}
\subsection{SEI growth}\label{SEI results}
\begin{figure}[h]
\centering
\includegraphics[width=0.8\linewidth]{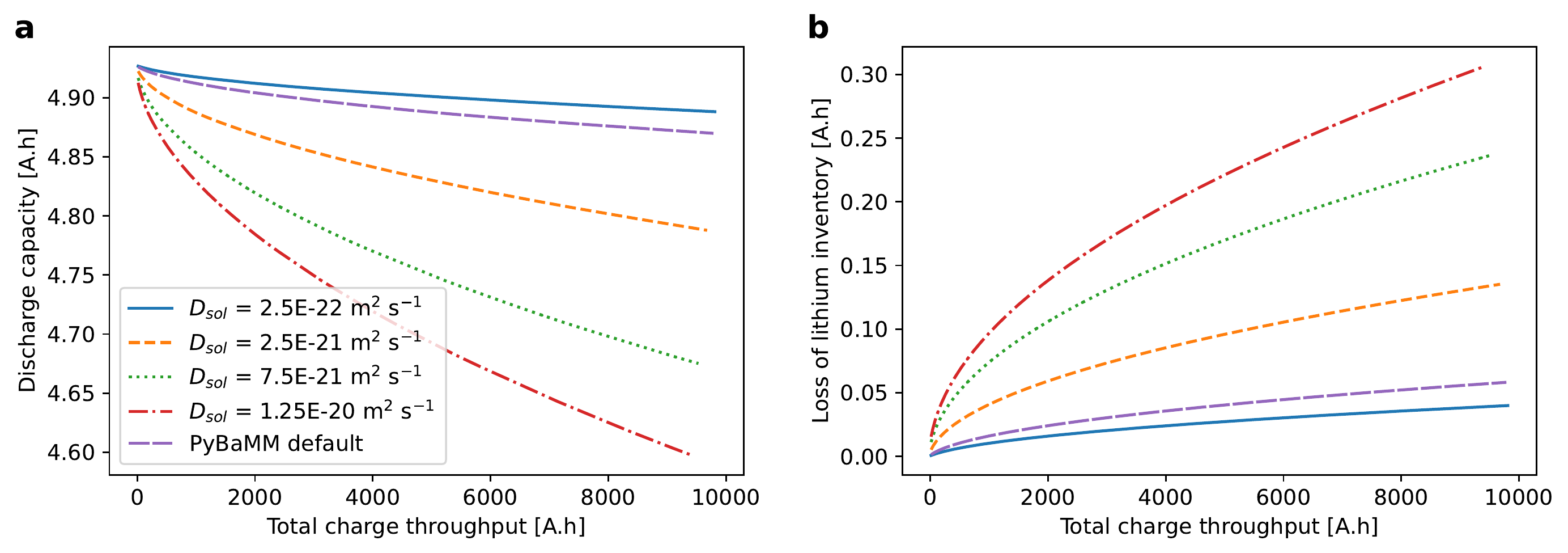}
\caption{Cell with SEI layer growth: (a) Decrease in discharge capacity over 1000 cycles at 25 $^\circ$C for different values of the SEI diffusivity $D_\text{sol}$; (b) loss of lithium inventory. PyBaMM's default two-layer SEI model is also included for comparison, with $D_\text{sol}=2.5\times10^{-22}$} m\textsuperscript{2} s\textsuperscript{-1}.
\label{SEI figure}
\end{figure}

In the first parametric study, the diffusion-limited SEI model outlined in section \ref{SEI model} is run with different values of the solvent diffusion coefficient $D_\text{sol}$ and the standard cycling protocol is applied. All degradation mechanisms other than SEI growth, i.e. Li plating, SEI on cracks and LAM, are disabled.\\\\The reduction in 1C discharge capacity over 1000 cycles is shown in Fig. \ref{SEI figure} (a), and is matched by the loss of lithium inventory plotted in Fig. \ref{SEI figure} (b). The rate of capacity fade gets slower over time as expected for diffusion-limited SEI growth \cite{Safari2009MultimodalBatteries,Reniers2019ReviewBatteries}.\\\\For the sake of simplicity when interacting with other degradation mechanisms, an SEI with only one layer is considered. By default, PyBaMM considers a two-layer SEI, so the results for two layers and PyBaMM's default value of $D_\text{sol}=2.5\times10^{-22}$ m\textsuperscript{2} s\textsuperscript{-1} are also plotted in Fig. \ref{SEI figure}. The effect of having two SEI layers instead of one is small compared to that of changing the solvent diffusivity by an order of magnitude.
\subsection{Li plating}
\begin{figure}[t]
\centering
\includegraphics[width=0.8\linewidth]{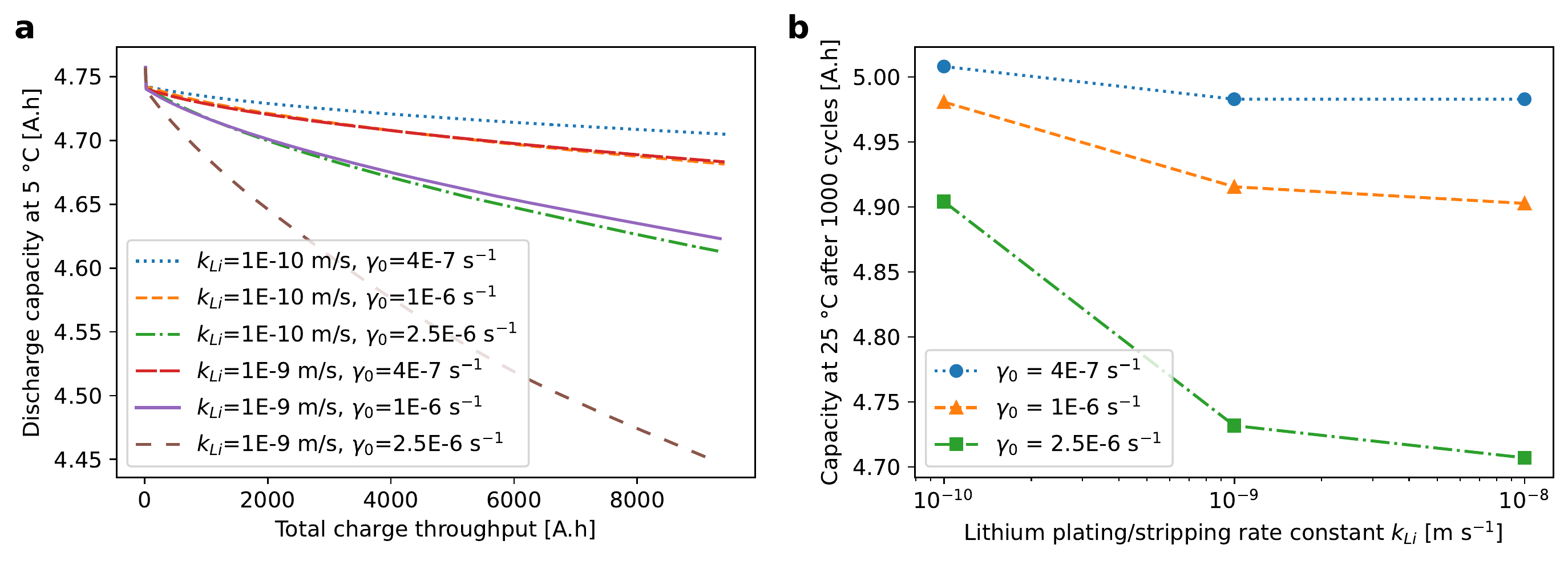}
\caption{(a) Decrease in cell capacity at 278.15 K (5 $^\circ$C) over 1000 cycles due to SEI growth and partially reversible Li plating. (b) Variation of capacity at 298.15 K (25 $^\circ$C) after 1000 cycles with plating/stripping constant $k_\mathrm{Li}$ and dead Li formation constant $\gamma_0$. Increasing $k_\mathrm{Li}$ beyond $10^{-9}$ m s\textsuperscript{-1} has little effect on capacity, while $\gamma_0$ always has a large impact.}
\label{plating figure}
\end{figure}

To investigate the effect of the new time-limited dead Li formation model proposed in section \ref{plating model} on capacity, both SEI growth and Li plating are enabled, while SEI on cracks and LAM are disabled. The standard cycling protocol is used, but the 1000 degradation cycles are done at the lower temperature of 5 $^\circ$C, in order to increase the effect of Li plating relative to SEI growth. Both of the characterization cycles are done at 25 $^\circ$C. The cell capacity at 5 $^\circ$C over 1000 cycles is plotted in Fig. \ref{plating figure} (a) for a range of values of the two unknown constants $k_\mathrm{Li}$ and the decay rate $\gamma_0$ for the formation of dead lithium. All capacity curves follow the same trend, because the decay of plated Li into dead Li is controlled by $\gamma_0 \div L_\text{SEI}$, and $L_\text{SEI}$ evolves in the same way regardless of plating in this model.\\\\
The effect of the magnitude of the two constants $k_\mathrm{Li}$ and $\gamma_0$ can be analysed by considering the cell capacity at 25 $^\circ$C evaluated from the final characterization cycle, plotted in Fig. \ref{plating figure} (b). For $k_\mathrm{Li}>10^{-9}$ m s\textsuperscript{-1}, the capacity fade is much more sensitive to $\gamma_0$ than $k_\mathrm{Li}$, while for $k_\mathrm{Li}<10^{-9}$ m s\textsuperscript{-1}, both constants have a large impact. This can be explained by the fact that increasing $k_\mathrm{Li}$ increases the rate of both plating and stripping; additional Li plated by increasing $k_\mathrm{Li}$ above $10^{-9}$ m s\textsuperscript{-1} is stripped before it decays into dead Li.
\subsection{Cracking and SEI}\label{crack results}
The particle cracking model outlined in section \ref{crack model} is combined with the SEI growth model outlined in Section \ref{SEI model}, with Li plating and LAM disabled. Degradation over 1000 cycles is tracked for the standard rate of particle cracking, as in Table S5, as well as for three accelerated cracking rates. All other parameters and the cycling conditions are per the default.
\begin{figure}[t]
	\centering
	\includegraphics[width=0.8\linewidth]{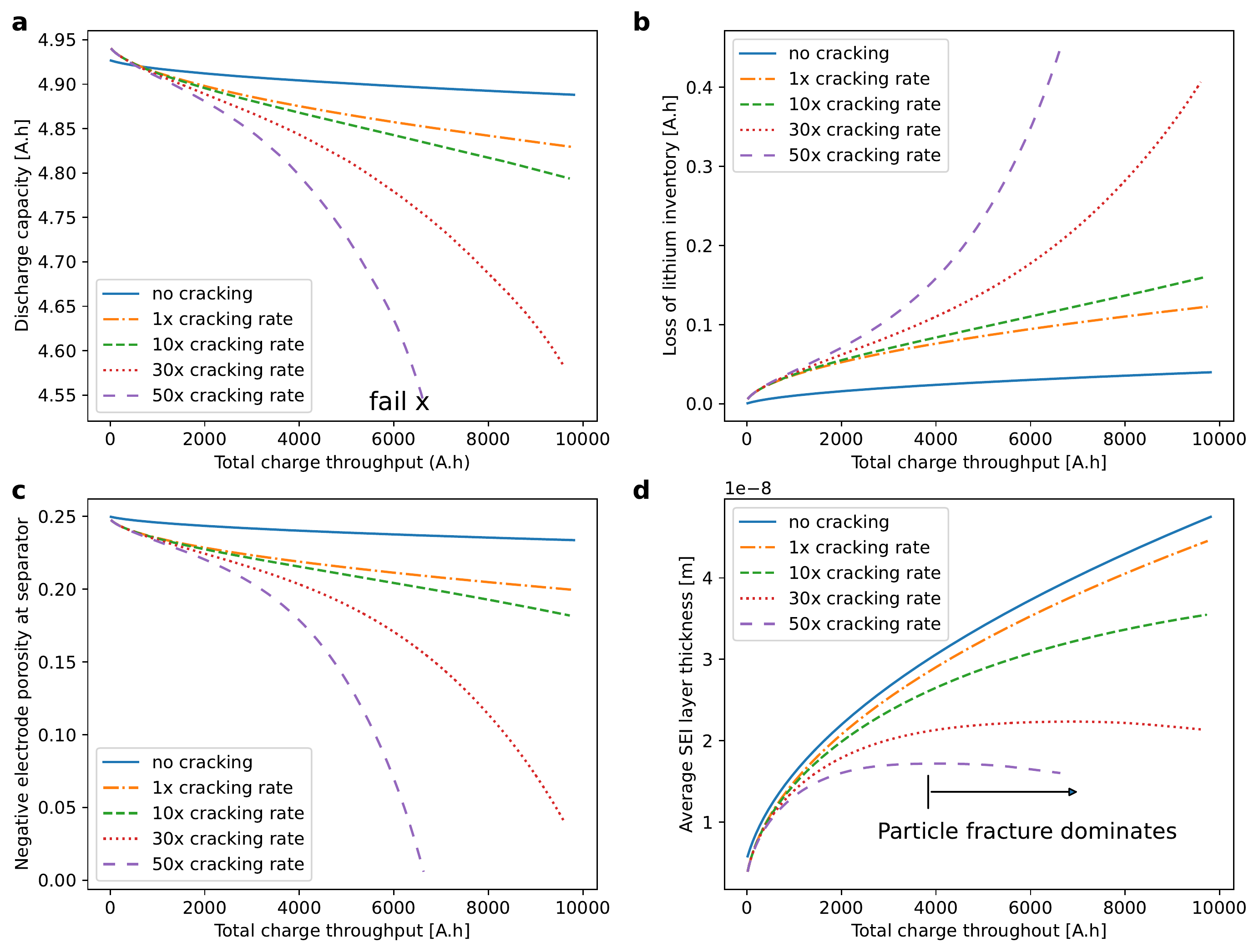}
	\caption{Standard cycling protocol with SEI and particle cracking enabled: (a) discharge capacity degradation, (b) loss of lithium inventory, (c) porosity reduction and (d) averaged SEI layer thickness.}
	\label{cracking figure} 
\end{figure}
\\\\The capacity fade is shown in Fig. \ref{cracking figure} (a). For the lowest cracking rate, degradation decreases with time, as expected for diffusion-limited SEI growth. For the 10x cracking rate, the capacity fade is quasi-linear. For the largest cracking rates, a knee point is visible after 5000 Ah and 3000 Ah throughput, for the 30x and the 50x cracking rate, respectively. For the 50x cracking rate, sudden failure occurs at 6600 Ah throughput.\\\\The loss of lithium inventory caused by the SEI is shown in Fig. \ref{cracking figure} (b). For low cracking rates, the LLI varies as the square root of time, a result of the solvent diffusion limited SEI growth model. For larger cracking rates, the LLI starts with a square root dependence, but then accelerates as the cracks propagate. The accelerated LLI is the cause for the knee point transition in the capacity loss. The additional surface exposed by particle cracking is shown in Fig. S2.\\\\The reason for the sudden failure in the 50x cracking rate is that the porosity at the negative electrode-separator interface reaches zero, as shown in Fig. \ref{cracking figure} (c). The cell with 30x cracking rate would fail in the same way if the simulation carried on after 1000 cycles.\\\\The case of particle cracking uses the assumption of microcracks, which is different from the ideal sphere assumption in the case of no cracking (also the standard Newman battery model). The averaged SEI layer thickness is presented in Fig. \ref{cracking figure} (d). For the cases of 30x and 50x cracking rate it stops increasing after 7500 Ah and 4000 Ah throughput respectively, indicating that the cracks are growing faster than the SEI can passivate them. This causes the runaway pore clogging shown in Fig. \ref{cracking figure} (c), leading to sudden failure. \cite{Garrick2017, Rahe2019}

\subsection{Loss of active material}\label{LAM results}
The loss of active material (LAM) due to particle cracking in the positive and negative electrodes is modelled as outlined in section \ref{LAM model}. The effects of LAM on cell degradation are studied while all other degradation mechanisms - SEI, lithium plating and SEI on cracks - are disabled. The governing equation \eqref{eq:LAM_t} indicates that the amount of LAM is dependent on the magnitude of the hydrostatic tensile stress; compressive stress make no contribution. Model predictions for cell degradation are run for five different combinations of the proportionality constant $\beta$. The negative and positive electrodes have independent rates, with ``n$x$p$y$" denoting the negative and positive electrode $\beta$ respectively. The values of $x$ and $y$ are proportionality constants applied to the reference $\beta$ values in Table S5.
\begin{figure}[t]
	\centering
	\includegraphics[width=\linewidth]{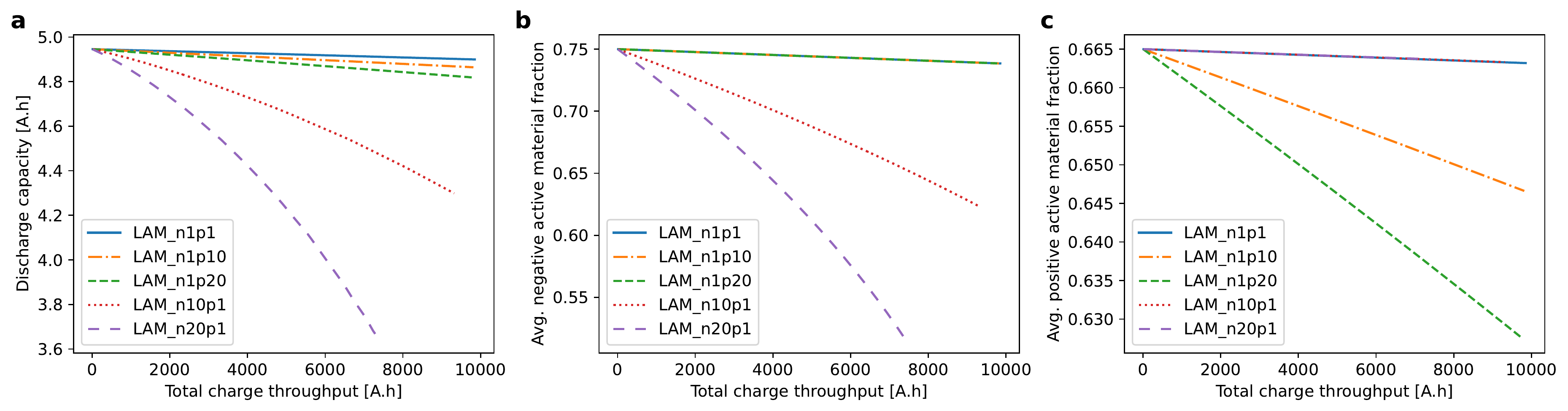}
	\caption{Model predictions for the effect of LAM due to particle cracking: (a) discharge capacity fade; averaged volume fraction of active materials in the (b) negative electrode and (c) positive electrode.}
	\label{LAM figure} 
\end{figure}
\\\\The discharge capacity fade, shown in Fig. \ref{LAM figure} (a), decreases quasi-linearly with cycle number for n1p1, n1p10 and n1p20, and non-linearly for the other two cases. The loss of active material in each electrode is shown in Fig. \eqref{LAM figure} (b) and (c). For instance, the active material fraction in the negative electrode decreases by 30\% for the case n20p1, corresponding to 25\% capacity loss. The capacity fade is stronger than linear because of the increase of the interfacial current density, as shown in Fig. S3, leading to higher stresses. For the same proportionality constant $\beta$, the LAM in the negative electrode is larger than that in the positive electrode, as visible, for example, between n1p20 and n20p1. This occurs for two reasons. Firstly, the discharge rate in the standard cycling protocol is 1C, whereas the charge rate is only 0.3C. Inspection of \eqref{sigma_r} and \eqref{sigma_t} shows that only the electrode from which Li\textsuperscript{+} de-intercalates experiences tensile (positive) stress, and only tensile stress contributes to LAM. Secondly, the positive electrode can be subjected to a higher tensile stress without cracking, as its critical stress is higher than that of the negative electrode, $\sigma_\mathrm{c}=3.75\times10^8$ Pa vs $6\times10^7$ Pa.

\subsection{All degradation mechanisms}
\begin{figure}[t]
    \centering
    \includegraphics[width=0.8\linewidth]{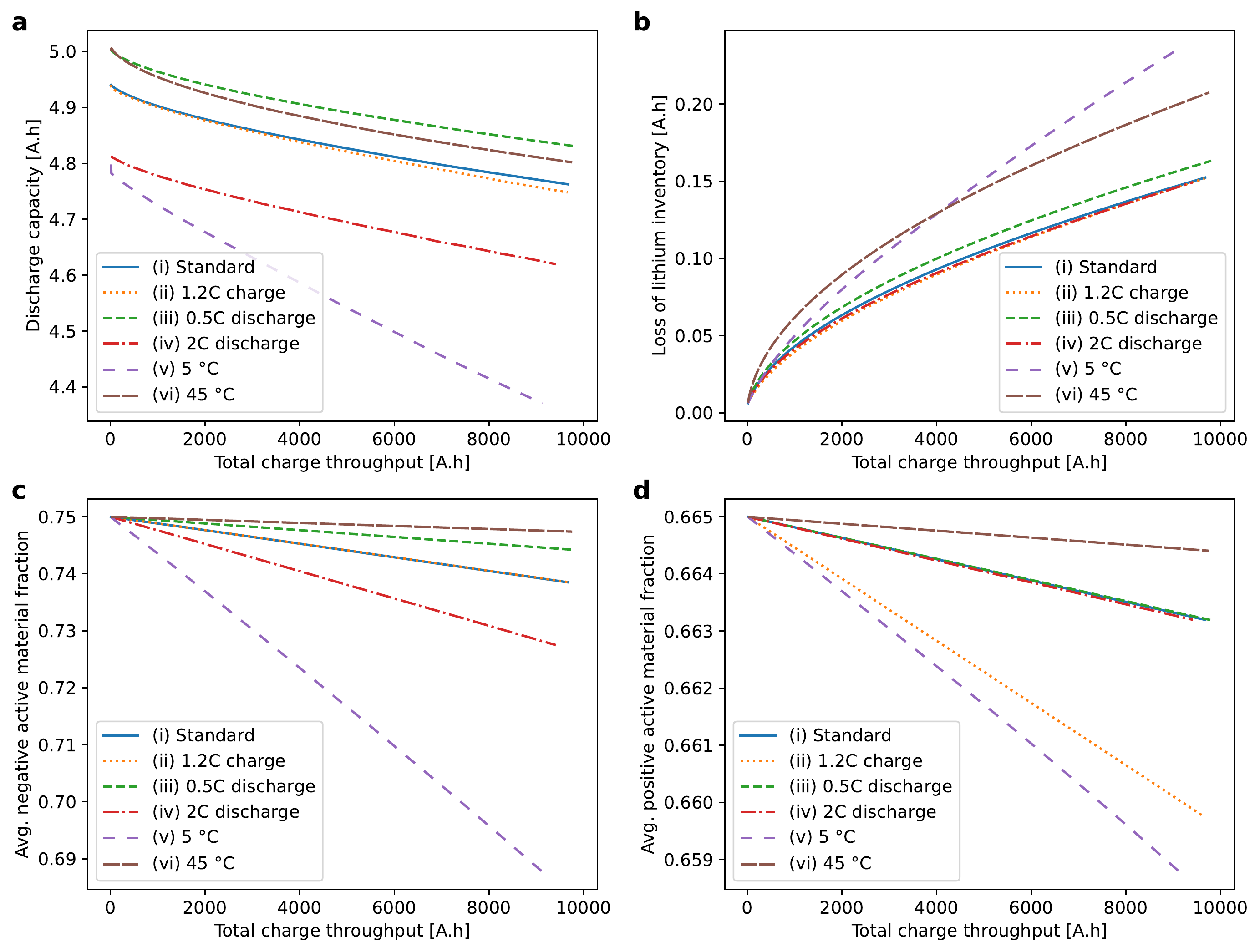}
        \caption{Variation of (a) discharge capacity, (b) loss of lithium inventory, (c) $x$-averaged negative active material fraction and (d) $x$-averaged positive active material fraction over 1000 cycles for the six cycling protocols in Table \ref{cycling protocols}.}
    \label{FullyCoupled}
\end{figure}

\begin{figure}[t]
    \centering
    \includegraphics[width=\linewidth]{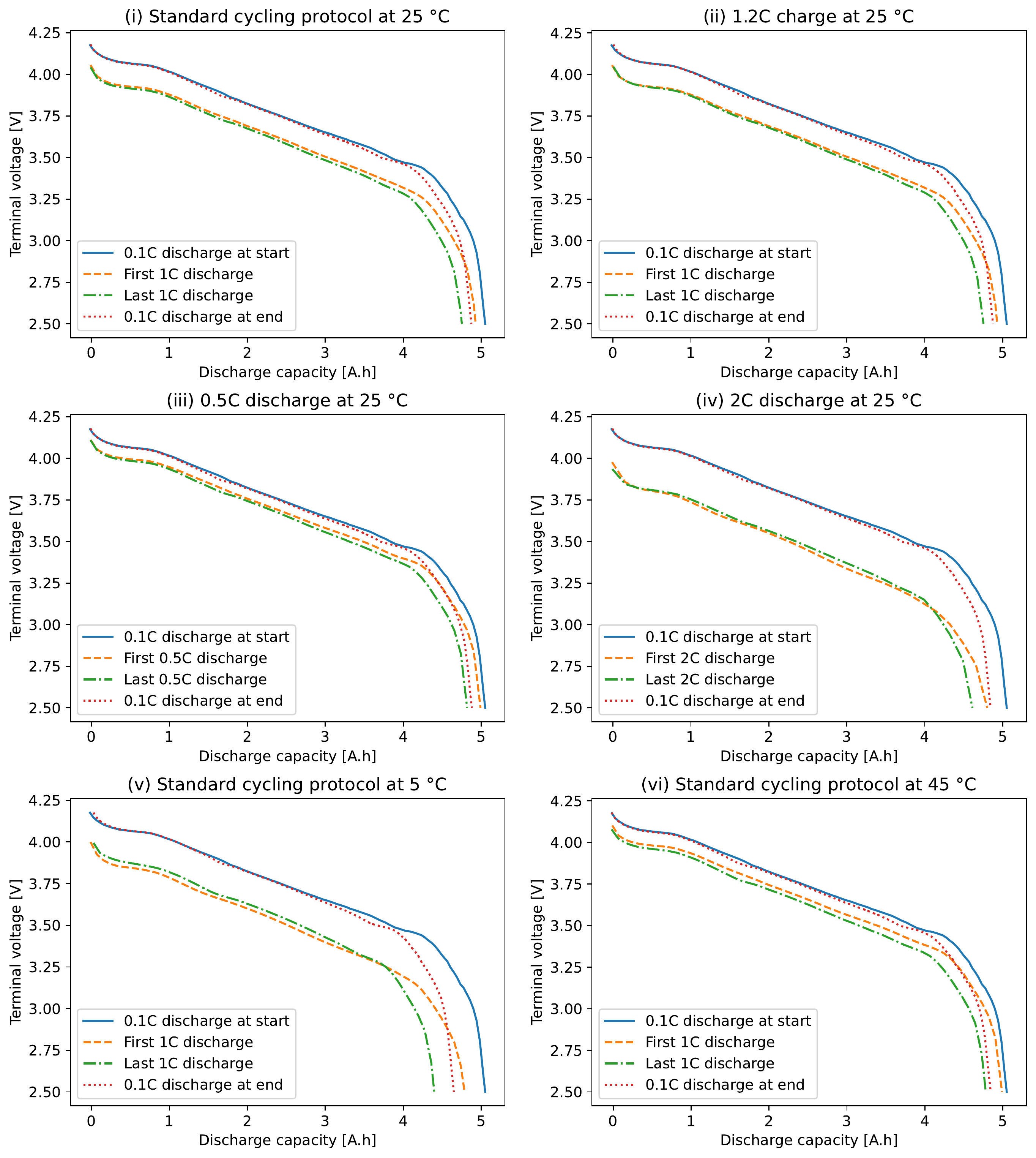}
    \caption{Cell voltage $V$ as a function of discharge capacity at four specific stages during cycling, for each of the six protocols in Table \ref{cycling protocols}: during the 0.1C discharge at 25~$^\circ$C before 1000 cycles, during the first cycle, during the last cycle and during the 0.1C discharge at 25~$^\circ$C after 1000 cycles.}
    \label{ZFLE}
\end{figure}

With all degradation mechanisms - SEI layer growth, lithium plating, SEI on cracks and LAM - enabled, model predictions are created using the standard cycling protocol and five variations, as described in Table \ref{cycling protocols}. The degradation parameters that were varied in Section \ref{SEI results}-\ref{LAM results} are assumed to have the default values listed in Table S5.
\begin{table}[h]
    \centering
    \caption{Cycling protocols used for the model with strong coupling between all degradation mechanisms: SEI layer growth, Li plating, SEI on cracks and LAM. The standard cycling protocol is denoted (i) and five variations of it are also considered.}
    \label{cycling protocols}
    \begin{tabular}{|c|c|c|c|}
         \hline Cycling protocol & Discharge rate & Charge rate & Temperature\\
         \hline (i) & 1C & 0.3C & 25 $^\circ$C\\
         \hline (ii) & 1C & 1.2C & 25 $^\circ$C\\
         \hline (iii) & 0.5C & 0.3C & 25 $^\circ$C\\
         \hline (iv) & 2C & 0.3C & 25 $^\circ$C\\
         \hline (v) & 1C & 0.3C & 5 $^\circ$C\\
         \hline (vi) & 1C & 0.3C & 45 $^\circ$C\\
         \hline
    \end{tabular}
\end{table}
\\\\The discharge capacity evolution over the 1000 cycles is plotted in Fig. \ref{FullyCoupled} (a) for each of the six protocols. The largest change in discharge capacity occurs for the discharge at 5 $^\circ$C, followed by the cell cycled at 45 $^\circ$C. The cells kept at 25 $^\circ$C degrade the least.\\\\
The total LLI from the three contributing mechanisms - surface SEI, SEI on cracks and dead lithium - is plotted in Fig. \ref{FullyCoupled} (b). For every cell except the one cycled at 5 $^\circ$C, the LLI accounts for most of the loss in discharge capacity. The relative contribution of each mechanism to LLI is depicted in Fig. S4.\\\\
The time evolution of the average active material volume fraction for each of the six protocols is shown in Fig. \ref{FullyCoupled} (c) for the negative electrode and in Fig. \ref{FullyCoupled} (d) for the positive electrode. Even for protocol (ii), in which the charge rate is faster than the discharge rate, more active material is lost in the negative electrode because the critical stress in the positive electrode is much larger. By far the greatest loss of active material occurs when the temperature is reduced to 5~$^\circ$C in protocol (v). Low temperatures reduce the solid diffusion coefficients $D_\pm$, increasing the Li\textsuperscript{+} concentration gradients within the electrode particles and therefore increasing mechanical stress.\\\\
However, the capacity plot in Fig. \ref{FullyCoupled} (a) only considers discharge capacity at 1C. Voltage-capacity plots for the `before and after' characterization cycles are compared with those of the first and last standard cycles in Fig. \ref{ZFLE}. In all cases, the capacity is different for all four of these cycles. However, the difference in capacity between the first and last 1C discharges is the same as that between the two corresponding 0.1C discharges, so the reduction in capacity due to degradation is not due to the increasing current. The difference in capacity between the first and last 0.1C characterization cycles is similar for all protocols except for the cell cycled at 5~$^\circ$C, which degraded much more than the others. This additional degradation is accounted for by the loss of active material as plotted in Fig. \ref{FullyCoupled} (c).

\subsection{Degradation paths}
\begin{figure}
    \centering
    \includegraphics[scale=0.64]{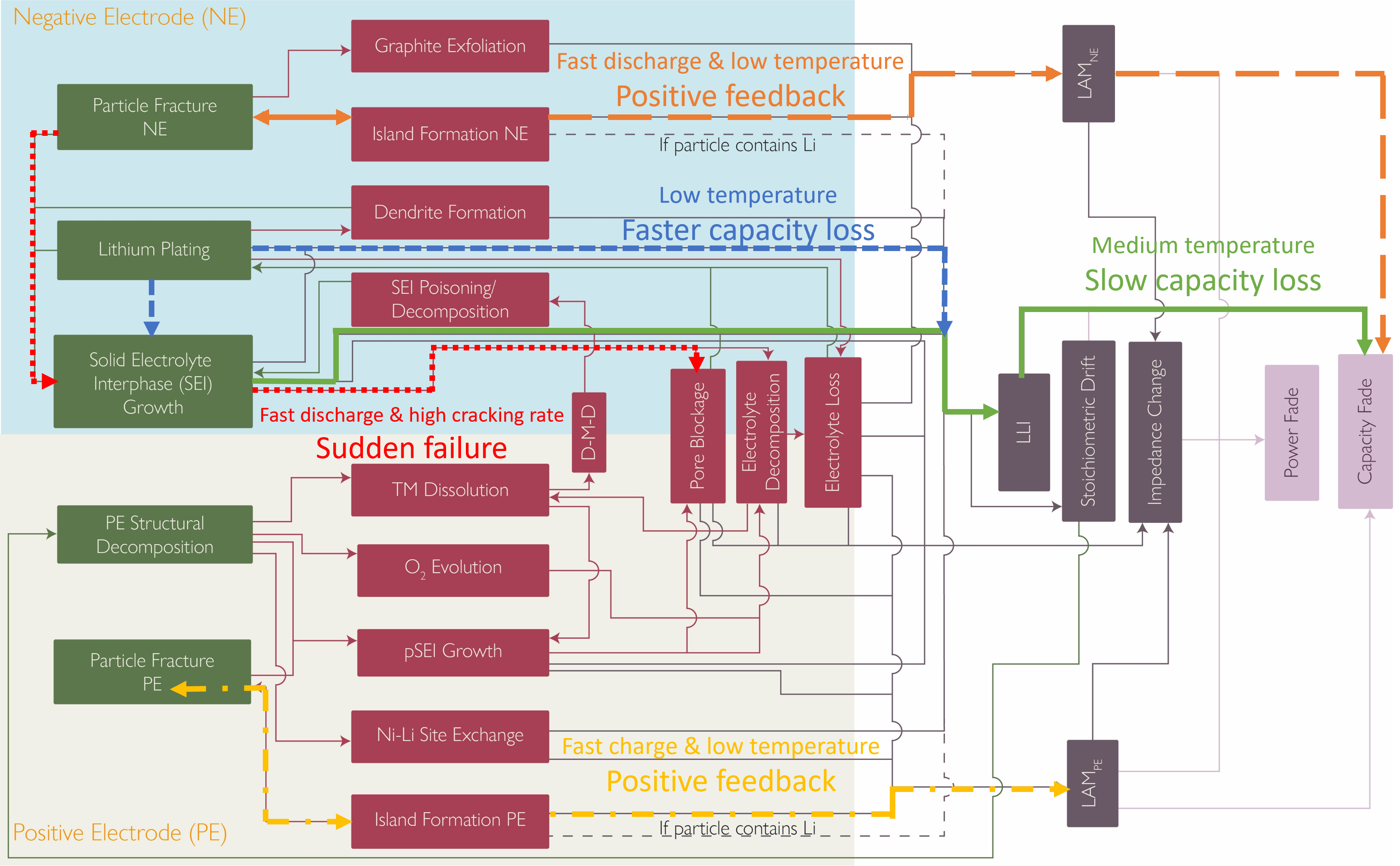}
    \caption{A summary of the interactions between degradation mechanisms that have been explored in this work. Adapted from Edge \etal \cite{Edge2021}}
    \label{degradation paths}
\end{figure}

The multitude of possible interactions between degradation mechanisms creates a maze of possible paths for a Li-ion battery, depicted in Fig. 9 of Edge \etal \cite{Edge2021}. Five of these paths have been identified in this work and have been overlaid on the original figure to create Fig. \ref{degradation paths}.\\\\The slowest degradation is achieved by cycling protocols (i)-(iv) in Table \ref{cycling protocols}, characterized by moderate temperature. In this case, the major degradation mechanism is a relatively slow SEI growth. Cycling protocol (vi), which occurs at a higher temperature, also follows this path but with a higher rate of degradation.\\\\
Cycling protocol (v) occurs at significantly lower temperatures. The dominant degradation path in this scenario is particle cracking causing LAM through island formation and binder delamination, resulting in higher interfacial current densities, increasing mechanical stress and thus accelerated LAM. The same self-reinforcing behaviour is achieved when the LAM parameter in the model is increased.\\\\Low temperatures also trigger faster capacity loss due to lithium plating, which speeds up capacity fade through the creation of dead lithium and additional SEI, the latter not included in this model. Unlike SEI growth, dead lithium formation is not self-inhibiting, so the resulting capacity fade is faster compared to scenarios with only SEI growth.\\\\
Figure \ref{FullyCoupled} (d) shows that decreasing the temperature or increasing the charge rate increases the loss of active material in the positive electrode. Either of these can activate the same positive feedback loop as in the negative electrode. However, the Young's modulus assumed for NMC 811 in this work was set too high for the loss of active material to make a significant contribution to overall capacity fade.\\\\
Only one of the scenarios modelled in this work reproduces the infamous ``cliff edge" or sudden failure, and it is displayed in Fig. \eqref{cracking figure}. When the Paris' Law cracking rate is increased, the cracks grow at a faster rate than the SEI can passivate them, resulting in runaway SEI growth that causes rapid pore clogging.\\\\As more degradation mechanisms are implemented and the parameter space is further explored, other scenarios where sudden failure occurs may be discovered. Importantly, even if sudden failure is predicted for a particular parameter set, the case is only relevant if that parameter set occurs in practice. While the sensitivity studies in this work are an important first step, it is equally important to carry out experimental work to find out what range of parameters is relevant for different cells and operating conditions.

\subsection{Recommendations for future model additions}\label{additions}
This work represents a significant step towards developing an all-encompassing model of degradation mechanisms in LiBs and harnessing the opportunities provided by the PyBaMM code. The main features recommended for future implementation are: adding thermal-electrochemical coupling, accounting for the effects of SiO\textsubscript{x} on the solid state diffusion, lithium plating and cell voltage hysteresis, allowing Li plating on fresh crack surfaces, allowing plated Li to transform into SEI, including a more comprehensive model of cathode degradation, and loss of electrolyte through solvent evaporation or consumption. Despite the strong dependence of lithium plating on nonlinear diffusion \cite{OKane2020}, it is not implemented here, because $D_\pm(c_s)$ is difficult to define for negative electrodes that contain both graphite and SiO\textsubscript{x} \cite{Chen2020a}. For negative electrodes with SiO\textsubscript{x}, the voltage hysteresis has recently been modelled \cite{Jiang2020VoltageAmorphization} and could be included.

\section{Conclusions}
Modelling of lithium-ion batteries is essential for the development of future electric vehicles and grid scale energy storage systems. Many modelling efforts have included degradation effects such as solid-electrolyte interphase growth, lithium plating, particle cracking and loss of active material. However, these were usually considered in isolation. There is an urgent need to develop understanding around the coupled nature of these degradation mechanisms and their effects. 
In this work, a model is developed in the PyBaMM software package that couples many of the relevant degradation mechanisms. Model predictions show that diffusion-limited SEI growth formula of Single \etal \cite{Single2018IdentifyingInterphase} results in the SEI thickness increasing with the square root of time, causing the loss of lithium inventory to increase with the same dependency. It is shown that increasing the solvent diffusivity by an order of magnitude has a larger effect than including a two-layer SEI.\\\\
The modified Butler-Volmer equation for Li plating/stripping pioneered by Yang \etal \cite{Yang2018ACells} is combined with a new model for a time-limited decay of plated Li to dead Li. For values of the Butler-Volmer rate constant larger than $10^{-9}$ m s\textsuperscript{-1}, the capacity fade becomes much more sensitive to the rate constant for dead Li formation than to the plating rate constant. For values of the Butler-Volmer rate constant smaller than $10^{-9}$ m s\textsuperscript{-1}, the capacity fade is highly sensitive to both parameters. The model therefore requires further parameterization work in order to be of any practical use.\\\\
When the particle cracking mechanism is included, the loss of lithium inventory increases significantly. For larger cracking rates, the porosity at the negative electrode-separator interface is shown to reach zero due to runaway SEI growth on cracks, causing sudden cell failure. This scenario demonstrates the increased predictive power of pseudo-2D models over single particle models, which track an average porosity only, and therefore would overestimate the battery's lifespan.\\\\
The formula for the loss of active material (LAM) proposed by Laresgoiti \etal \cite{Laresgoiti2015}, which works by calculating the mechanical stress in electrode particles, results in a positive feedback loop where degradation accelerates over time. The loss of active material has a larger effect in the negative electrode than in the positive electrode because the critical stress in the negative electrode is assumed to be over six times lower. This is therefore another degradation mechanism for which more accurate knowledge of key parameters is required.\\\\
When all four mechanisms are enabled, the model predicts low temperatures to cause the most irreversible capacity fade. While significant lithium plating occurs, the main cause of the observed capacity fade is due to loss of active material, which is greatly accelerated at low temperatures due to mechanical cracking of the negative electrode. The sudden cell failure due to pore clogging is also due to mechanical cracking. Therefore, the model predicts that avoiding cracking is just as important as avoiding lithium plating, if not more so.\\\\

\section*{Conflicts of interest}
The authors have no conflicts of interest to report.

\section*{Acknowledgement}
The authors would like to acknowledge financial support from EPSRC Faraday Institution Multiscale Modelling project (EP/S003053/1, grant number FIRG025).

\section*{Credit statement}
Simon E. J. O'Kane wrote the code for the lithium plating model, ran all simulations used to produce the final results and was involved in writing all sections of the manuscript. Weilong Ai wrote the code for the particle cracking and active material models and wrote the first drafts of sections \eqref{fracture}, \eqref{crack model}, \eqref{LAM model}, \eqref{pybamm}, \eqref{crack results} and \eqref{LAM results}. Ganesh Madabattula wrote a Jupyter notebook that was later adapted to produce the final results. Diego Alonso Albarez provided technical support with PyBaMM, managed the authors' interactions with the PyBaMM team and managed a GitHub repository that allowed the authors to share code. Robert Timms and Valentin Sulzer added the constant voltage charge protocol and event state feature to PyBaMM and also provided technical support with PyBaMM. Jacqueline Sophie Edge wrote the first draft of Section \eqref{additions} and provided strategic advice on many aspects of the work. Billy Wu provided many helpful comments on early drafts of the manuscript and supervised Weilong Ai's work on the mechanical model. Gregory J. Offer wrote the first draft of the abstract and Broader Context, while also providing the leadership to make this collaboration possible. Monica Marinescu carried out edits to all sections of the final version of the manuscript and supervised Simon E. J. O'Kane's development of the plating model and final PybaMM simulations.

\bibliographystyle{unsrt}
\bibliography{references}
\end{document}


\maketitle

\section*{DFN model equations}
The basic DFN model equations are listed in Table S1. The notation is the same as that used by O'Kane \etal \cite{OKane2020}, with two changes: $N_\mathrm{sr}$ now denotes side reactions in general as opposed to Li plating specifically, and the voltage drop $\eta_\mathrm{SEI}$ due to the SEI resistance has been added to the Butler-Volmer equation. 
 \begin{table}[h]
     \centering
     \renewcommand{\arraystretch}{2}
     \begin{tabular}{cc}
        \hline Variable & Equation \\
        \hline $\phi_\mathrm{s}(x,t)$ & $\sigma_\pm\dfrac{\partial^2\phi_\mathrm{s}}{\partial x^2}=j_\mathrm{tot}$\\
        $\phi_\mathrm{e}(x,t)$ & $-\kappa_\mathrm{eff}(c_\mathrm{e},T)\dfrac{\partial^2\phi_\mathrm{e}}{\partial x^2}+\dfrac{2RT}{F}\kappa_\mathrm{eff}(c_\mathrm{e},T)(1-t^+)\dfrac{\partial^2\ln c_\mathrm{e}}{\partial x^2}=j_\mathrm{tot}$\\
         $c_\mathrm{e}(x,t)$ & $\dfrac{\partial(\epsilon c_\mathrm{e})}{\partial t}=\dfrac{\partial}{\partial x}\left(D_\mathrm{eff}(c_\mathrm{e},T)\dfrac{\partial c_\mathrm{e}}{\partial x}\right)+\left(\dfrac{1-t^+}{F}\right)j_\mathrm{tot}$\\
         $c_\mathrm{a}(x,r,t)$ & $\dfrac{\partial c_\mathrm{a}}{\partial t}=\dfrac{1}{r^2}\dfrac{\partial}{\partial r}\left(D_\pm(T) r^2\dfrac{\partial c_\mathrm{a}}{\partial r}\right)$\\
         $c_\mathrm{s}(x,t)$ & $\dfrac{\partial c_s}{\partial r} = -\dfrac{N_\mathrm{int}}{D_\pm(T)}$\\
         $j_\mathrm{tot}$ & $j_\mathrm{tot}=Fa_\pm\left(N_\mathrm{int}+\sum N_\mathrm{sr}\right)$\\
         $N_\mathrm{int}$ & $N_{\mathrm{int}}=2k_\pm(T)\sqrt{\dfrac{c_\mathrm{e}}{c_\mathrm{eq}}(c_{\mathrm{m}\pm}-c_\mathrm{s})c_\mathrm{s}}\sinh\left(\dfrac{F\eta}{2RT}\right)$\\
         $\eta$ & $\eta=\phi_\mathrm{s}-\phi_\mathrm{e}-U_\pm(c_\mathrm{s})-\eta_\mathrm{SEI}$\\
         \hline
     \end{tabular}
     \caption{Equations of the Doyle-Fuller-Newman (DFN) model.}
     \label{DFN}
 \end{table}
\section*{Electrode parameters}
\begin{figure}
\centering
\includegraphics[scale=0.55]{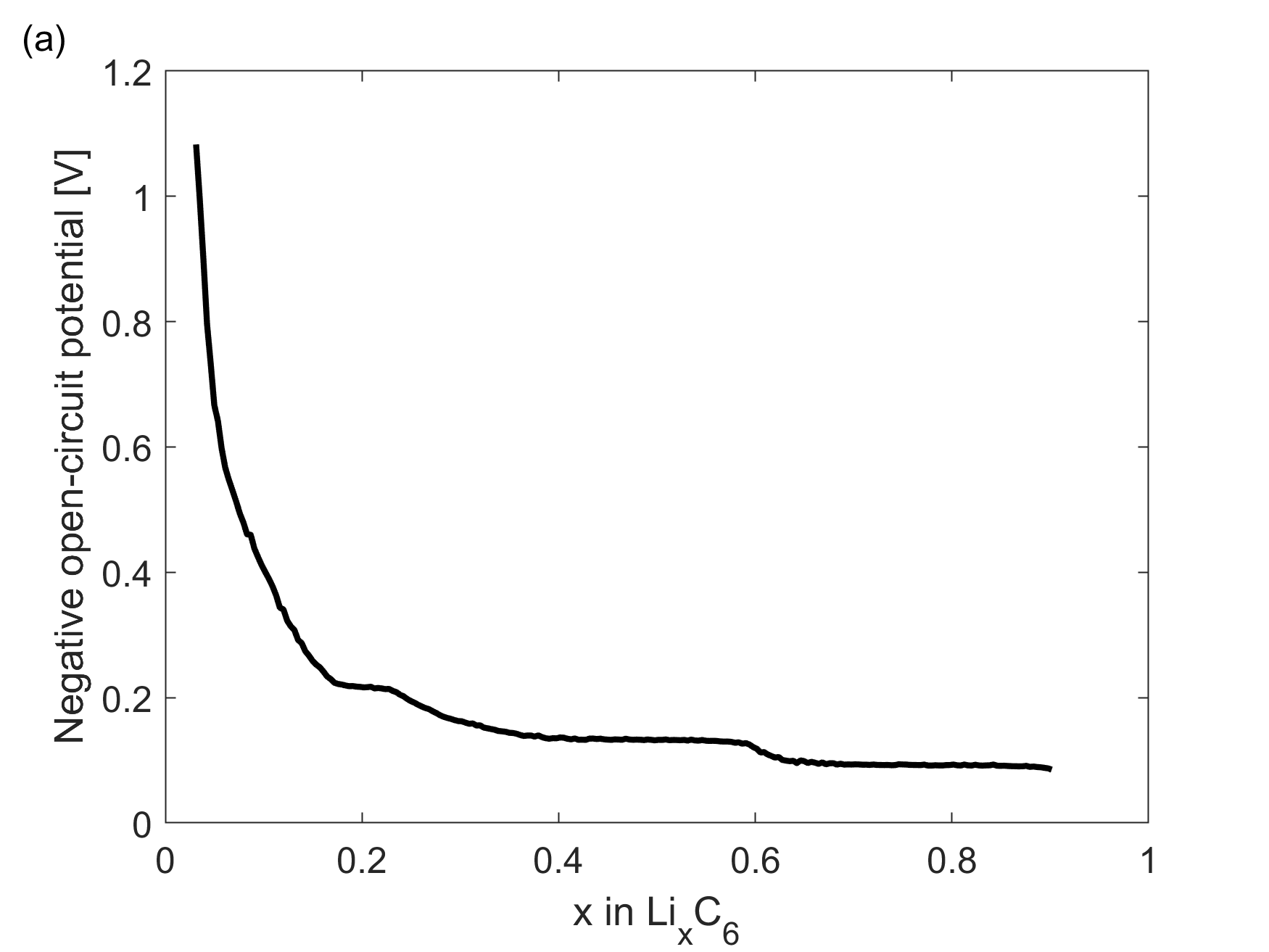}\includegraphics[scale=0.55]{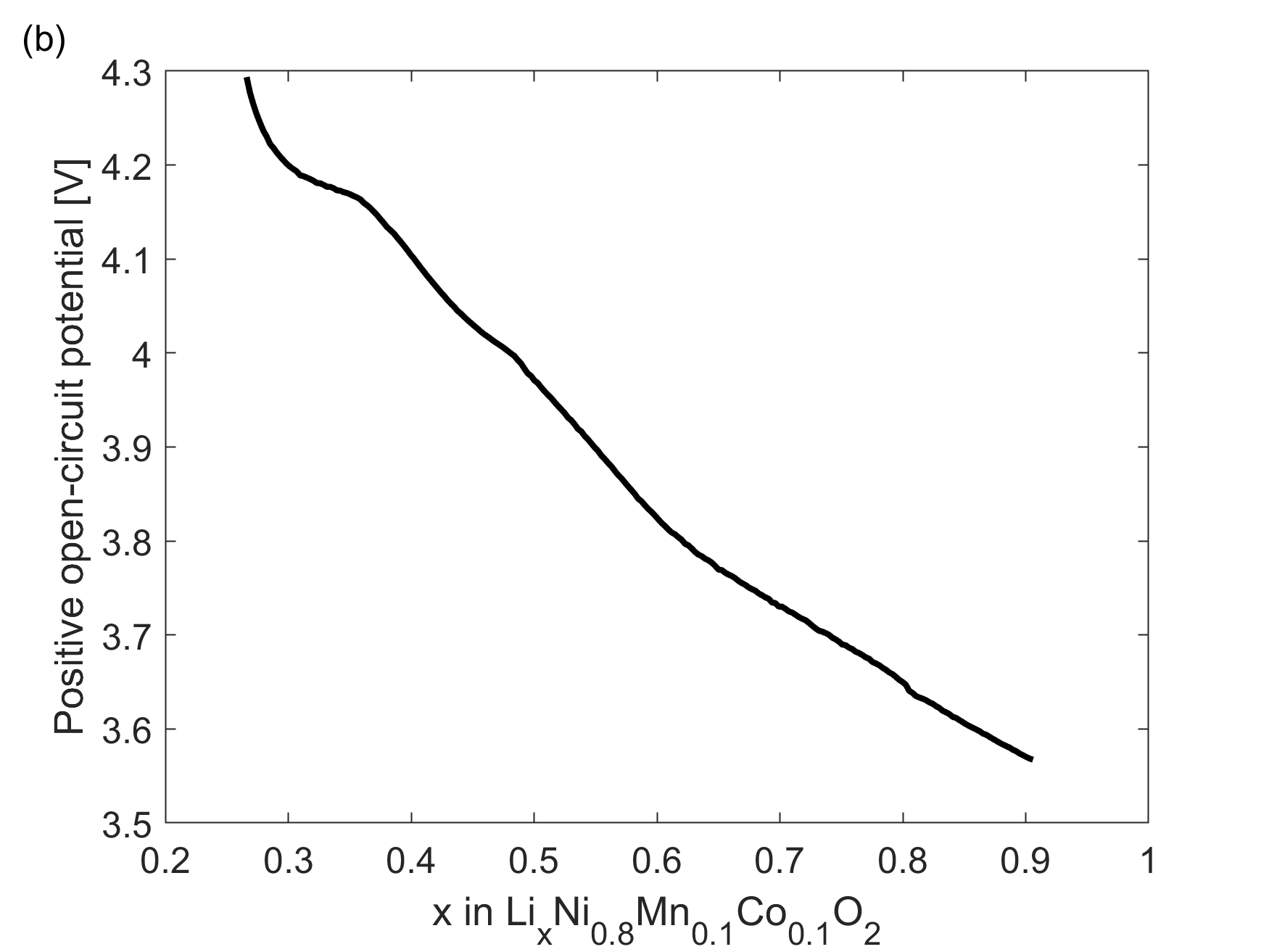}
\caption{Open-circuit potential $U_-(c_\mathrm{s}^*)$ of the (a) graphite+SiO\textsubscript{x} negative electrode and (b) NMC 811 positive electrode, as a function of normalized Li\textsuperscript{+} concentration, as measured by Chen \etal \cite{Chen2020a} at 25 $^\circ$C.}
\label{electrode graphs}
\end{figure}

The open-circuit potential curves $U_\pm(c_\mathrm{s}^*$) were measured by Chen \etal \cite{Chen2020a} at 25 $^\circ$C and are replotted in Fig. \ref{electrode graphs}. For both electrodes, the three-electrode cell measurements were used. Chen \etal found that the graphite+SiO\textsubscript{x} negative electrode showed significant hysteresis; the discharge branch of the OCP is used in the model, as in Chen \etal's own PyBaMM model.\\\\Other electrode parameters were taken from Tables VII and IX of Chen \etal \cite{Chen2020a} and are shown in Table \ref{electrode table}. The solid-state diffusion coefficients $D_\pm$ deserve special attention because O'Kane \etal \cite{OKane2020} identified $D_-$ as a critical parameter for Li plating/stripping. Chen \etal \cite{Chen2020a} made detailed measurements of $D_\pm$ as functions of Li\textsuperscript{+} concentration, but neither Chen \etal nor the authors of this work were able to implement this in PyBaMM. Instead, the negative electrode diffusivity $D_-$ is treated as a function of temperature only.\\\\
The temperature-dependent parameters $D_\pm(c_\mathrm{a}^*,T)$ and $k_\pm(T)$ are assumed to have Arrhenius temperature dependence:
\begin{gather}
D_\pm(T)=D_\pm(T_\mathrm{meas})\exp\left(\frac{E_{D\pm}}{RT_\mathrm{meas}}-\frac{E_{D\pm}}{RT}\right)\\
k_\pm(T)=k_\pm(T_\mathrm{meas})\exp\left(\frac{E_{k\pm}}{RT_\mathrm{meas}}-\frac{E_{k\pm}}{RT}\right),
\end{gather}
where $E_{D\pm}$ and $E_{k\pm}$ are activation energies and $T_\mathrm{meas}$ is the temperature at which detailed measurements were carried out, in this case 298.15 K (25 $^\circ$C).\\\\
However, Chen \etal \cite{Chen2020a} did not report temperature-dependent diffusivity data. For the negative electrode, an Arrhenius temperature dependence is assumed; the activation energy 30300 J mol\textsuperscript{-1} is taken from Ecker \etal \cite{Ecker2015} and is within one significant figure of two other values reported in the literature \cite{Kulova2006TemperatureGraphite, Schmalstieg2018a}. The authors are unaware of any temperature-dependent diffusivity data for NMC 811, but Caba\~nero \etal \cite{Cabanero2018} reported an activation energy of 25000 J mol\textsuperscript{-1} for the similar NCA material, so this value is taken.
\begin{table}
\centering
\def\arraystretch{1.2}
\begin{tabular}{cccc}
\hline {\bf Symbol} & {\bf Definition} & {\bf - electrode} & {\bf + electrode}\\
\hline $A$ & Total planar electrode area, m\textsuperscript{2} & 0.1027 & 0.1027\\
$a_\pm$ & Surface area to volume ratio, m\textsuperscript{-1} & $3.84\times10^5$ & $3.82\times10^5$\\
$c_{\mathrm{m}\pm}$ & Maximum Li\textsuperscript{+} concentration, mol m\textsuperscript{-3} & 33133 & 63104\\
$c_{0\pm}$ & Initial Li\textsuperscript{+} concentration, mol m\textsuperscript{-3} & 29866 & 17038\\
$D_\pm$ & Li\textsuperscript{+} diffusion coefficient at 25 $^\circ$C, m\textsuperscript{2} s\textsuperscript{-1} & $3.3\times10^{-14}$ & $4\times10^{-15}$\\
$E_{D\pm}$ & Activation energy for Li\textsuperscript{+} diffusion, J mol\textsuperscript{-1} & 30300\cite{Ecker2015} & 25000\cite{Cabanero2018}\\
$E_{k\pm}$ & Activation energy for rate constant, J mol\textsuperscript{-1} & 35000 & 17800\\
$k_\pm$ & (De)intercalation rate constant at 25 $^\circ$C, m s\textsuperscript{-1} & $2.12\times10^{-10}$ & $1.12\times10^{-9}$\\
$r_\pm$ & Electrode particle radius, m & $5.86\times10^{-6}$ & $5.22\times10^{-6}$\\
$\delta_\pm$ & Electrode thickness, m & $8.52\times10^{-5}$ & $7.56\times10^{-5}$\\
$\epsilon_\mathrm{e}$ & Electrolyte volume fraction & 0.25 & 0.335\\
$\epsilon_\mathrm{a}$ & Active material volume fraction & 0.75 & 0.665\\
$\sigma_\pm$ & Electrode conductivity, S m\textsuperscript{-1} & 215 & 0.18\\
\hline
\end{tabular}
\caption{Electrode parameters for the beginning of life model. All values taken from the final model values from Tables VII and IX in Chen \etal \cite{Chen2020a} unless otherwise specified.}
\label{electrode table}
\end{table}
\section*{Electrolyte parameters}
The effective conductivity $\kappa_\mathrm{eff}(c_\mathrm{e},T)$ and diffusion coefficient $D_\mathrm{eff}(c_\mathrm{e},T)$ of electrolyte occupying volume fraction $\epsilon$ are related to the corresponding values $\kappa(c_\mathrm{e},T)$ and $D_\mathrm{e}(c_\mathrm{e},T)$ in pure electrolyte by
\begin{equation}
\kappa_\mathrm{eff}(c_\mathrm{e},T)=\epsilon_\mathrm{e}^{1.5}\kappa(c_\mathrm{e},T)\quad\text{and}\quad D_\mathrm{eff}(c_\mathrm{e},T)=\epsilon_\mathrm{e}^{1.5}D_\mathrm{e}(c_\mathrm{e},T).
\end{equation}
Both $\kappa(c_\mathrm{e},T)$ and $D_\mathrm{e}(c_\mathrm{e},T)$ have an Arrhenius temperature dependence:
\begin{gather}
\kappa(c_\mathrm{e},T)=\kappa(c_\mathrm{e},T_\mathrm{meas})\exp\left(\frac{E_\kappa}{RT_\mathrm{meas}}-\frac{E_{\kappa}}{RT}\right)\\D_\mathrm{e}(c_\mathrm{e},T)=D_\mathrm{e}(c_\mathrm{e},T_\mathrm{meas})\exp\left(\frac{E_\kappa}{RT_\mathrm{meas}}-\frac{E_{\kappa}}{RT}\right),
\end{gather}
where $E_\kappa$ is the activation energy for both $\kappa$ and $D_\mathrm{e}$, $T_\mathrm{meas}$ is the temperature at which detailed measurements were carried out (in this case, 298.15 K), $\kappa(c_\mathrm{e},T_\mathrm{meas})$ is a cubic polynomial \cite{Chen2020a}
\begin{equation}\label{kappa}
\kappa(c_\mathrm{e},T_\mathrm{meas})=1.297\times10^{-10}c_\mathrm{e}^3-7.94\times10^{-5}c_\mathrm{e}^{1.5}+3.329\times10^{-3}c_\mathrm{e}
\end{equation}
and $D_\mathrm{e}(c_\mathrm{e},T)$ is a quadratic polynomial \cite{Chen2020a}
\begin{equation}\label{D_e}
D_\mathrm{e}(c_\mathrm{e},T_\mathrm{meas})=8.794\times10^{-17}c_\mathrm{e}^2-3.972\times10^{-13}c_\mathrm{e}+4.862\times10^{-10}.
\end{equation}
In \eqref{kappa} and \eqref{D_e}, $\kappa$ has units of S m\textsuperscript{-1}, $D_\mathrm{e}(c_\mathrm{e},T)$ has units of m\textsuperscript{2} s\textsuperscript{-1} and $c_\mathrm{e}$ has units of mol m\textsuperscript{-3}. The remaining parameters are taken from Table VII of Chen \etal \cite{Chen2020a} and listed in Table \ref{the other table}.
\begin{table}
\centering
\def\arraystretch{1.2}
\begin{tabular}{ccc}
\hline {\bf Symbol} & {\bf Definition} & {\bf Value}\\
\hline $c_\mathrm{eq}$ & Equilibrium Li\textsuperscript{+} concentration in electrolyte, mol m\textsuperscript{-3} & 1000\\
$E_\kappa$ & Activation energy for electrolyte conductivity, J mol\textsuperscript{-1} & 17100\cite{Ecker2015}\\
$F$ & Faraday's constant, C mol\textsuperscript{-1} & 96485\\
$Q_\mathrm{nom}$ & Nominal capacity, mAh & 5000\\
$R$ & Universal gas constant, J K\textsuperscript{-1} mol\textsuperscript{-1} & 8.314\\
$t^+$ & Li\textsuperscript{+} transference number & 0.2594\\
$V_\mathrm{max}$ & Upper cutoff voltage, V & 4.2\\
$V_\mathrm{min}$ & Lower cutoff voltage, V & 2.5\\
$\delta_\mathrm{s}$ & Separator thickness, m & $1.2\times10^{-5}$\\
$\epsilon_\mathrm{e}$ & Separator porosity & 0.47\\
\hline
\end{tabular}
\caption{Other parameters used in the model. All values taken from Chen \etal \cite{Chen2020a}}
\label{the other table}
\end{table}
\section*{Degradation parameters}
The parameters concerning battery degradation were not measured by Chen \etal \cite{Chen2020a}. The default degradation parameters in PyBaMM are taken from a range of sources and listed in Table \ref{tb:degradation parameters}. Some degradation parameters were varied as part of parametric studies and are assumed to have the values listed in Table \ref{tb:variable parameters} except in the study where that parameter is varied.

 \begin{table}[h]
 	\small
     \centering
     \renewcommand{\arraystretch}{1.5}
      \caption{Degradation parameters used in the model, except for those that were varied during the parametric studies, which are given in Table \ref{tb:variable parameters}.}
     \label{tb:degradation parameters}
     \begin{tabular}{cccccc}
     \hline 
     & & \multicolumn{2}{c}{Negative electrode}  & \multicolumn{2}{c}{Positive electrode}\\
     Symbol & Definition & Value & Ref. & Value & Ref.\\
     \hline $c_\mathrm{sol,0}$ & Bulk solvent concentration, mol m\textsuperscript{-3} & 2636 & \cite{Ploehn2004SolventCells}&&\\
      $\bar{V}_\mathrm{SEI}$ & SEI partial molar volume, m\textsuperscript{3} mol\textsuperscript{-1} & $9.585\times10^{-5}$ & \cite{Safari2009MultimodalBatteries}&&\\
      $\rho_\mathrm{SEI}$ & SEI resistivity, $\Omega$ m & $2\times10^5$ & \cite{Safari2009MultimodalBatteries}&&\\
      $L_\mathrm{SEI,0}$ & Initial SEI thickness, m & $5\times10^{-9}$ & \cite{Safari2009MultimodalBatteries}&&\\
      $E_\text{sol}$ & Solvent diffusion activation energy, J mol\textsuperscript{-1} & 37000 & \cite{Waldmann2014}\\
      $\alpha_\mathrm{a,Li}$ & Anodic transfer coefficient for Li stripping & 0.35 & Assumed\\
      $\alpha_\mathrm{c,Li}$ & Cathodic transfer coefficient for Li plating & 0.65 & Assumed\\
      $E$ & Young's modulus [Pa] & $1.5\times10^{10}$ & \cite{Ai2020ElectrochemicalCells} & $3.75\times10^{11}$ & \cite{Ai2020ElectrochemicalCells} \\
      $\nu$ & Poisson's ratio &  0.3  & \cite{Ai2020ElectrochemicalCells} & 0.2 & \cite{Ai2020ElectrochemicalCells} \\
      $\Omega$ & Partial molar volume [m$^3$/mol] & $3.1\times10^{-6}$  & \cite{Ai2020ElectrochemicalCells} &  $1.25\times10^{-5}$&\cite{Xu2019}   \\
      $l_\text{cr,0}$ & Initial crack length [m] & $2\times10^{-5}$ & \cite{Purewal2014a} & $2\times10^{-5}$ & \cite{Purewal2014a}   \\
      $w_\text{cr}$ & Initial crack width [m] & $1.5\times10^{-5}$ & \cite{Purewal2014a}  & $1.5\times10^{-5}$ & \cite{Purewal2014a} \\
      $\rho_\text{cr}$ & number of cracks per unit area [m$^{-2}$] & $3.18\times10^{15}$& \cite{Purewal2014a} & $3.18\times10^{15}$ & \cite{Purewal2014a} \\
      $b_\text{cr}$ & Stress intensity factor correction & 1.12 & \cite{Purewal2014a} & 1.12 & \cite{Purewal2014a} \\
	
      $m_\text{cr}$ & Paris' law exponential term & 2.2 & \cite{Purewal2014a} & 2.2 & \cite{Purewal2014a} \\	 
      		$\sigma_\text{c}$ & Critical stress for particle fracture [Pa] & $6\times10^7$ & Assumed & $3.75\times10^8$ & Assumed \\
      $m_2$ & Loss of active material exponential term & 2 & Assumed & 2 & Assumed  \\
     \hline 
     \end{tabular}

 \end{table}
 \begin{table}[h]
 	\small
     \centering
     \renewcommand{\arraystretch}{1.5}
      \caption{Degradation parameters varied during the parametric studies, along with their default values.}
     \label{tb:variable parameters}
     \begin{tabular}{cccccc}
     	     \hline 
     	& & \multicolumn{2}{c}{Negative electrode}  & \multicolumn{2}{c}{Positive electrode}\\
      Symbol & Definition & Default value & Ref. &  Default value & Ref.\\
     \hline $D_\mathrm{sol}$ & Solvent diffusivity in SEI, m\textsuperscript{2} s\textsuperscript{-1} & $2.5\times10^{-22}$ & \cite{Single2018IdentifyingInterphase}\\
      $k_\mathrm{Li}$ & Li plating/stripping rate constant, m s\textsuperscript{-1} & $10^{-9}$ & Assumed\\
      $\gamma_0$ & Rate constant for dead Li formation, s\textsuperscript{-1} & $10^{-6}$ & Assumed\\
            $k_\text{cr}$ & Paris' law cracking rate & $3.9\times10^{-20}$ & \cite{Purewal2014a} & $3.9\times10^{-20}$ & \cite{Purewal2014a} \\
     $\beta$ & Loss of active material proportional term & 0.001 & Assumed & 0.001 & Assumed\\	
     \hline 
     \end{tabular}

 \end{table}

\section*{Two-layer diffusion-limited SEI growth model}
In two-layer SEI models, the SEI thickness $L_\text{SEI}$ is replaced with two thicknesses $L_\text{inner}$ and $L_\text{outer}$ for the inner and outer layers respectively. It is assumed that the solvent can only diffuse through the outer layer, so the boundary conditions on the solvent concentration $c_\text{sol}$ become
\begin{gather}
    N_\text{sol}=-D_\text{sol}(T)\frac{\partial c_\text{sol}}{\partial l},\\
    c_\text{sol} =0 \quad \text{at} \quad l=L_\text{inner},\\
    c_\text{sol}=c_\text{sol,0} \quad \text{at} \quad l=L_\text{outer},
\end{gather}
The solution is
\begin{gather}
    c_\text{sol}=\frac{lc_\text{sol,0}}{L_\text{outer}},\\
    N_\text{sol}=-\frac{c_\text{sol,0} D_\text{sol}(T)}{L_\text{outer}},\\
    N_\text{inner} = -\tfrac{1}{2}N_\text{sol}=\frac{c_\text{sol,0} D_\text{sol}(T)}{L_\text{outer}},\\
    N_\text{outer} = -\tfrac{1}{2}N_\text{sol}=\frac{c_\text{sol,0} D_\text{sol}(T)}{L_\text{outer}},\\
\end{gather}
assuming the two layers grow at the same rate. Two differential equations are required, one for each layer: 
\begin{gather}
    \frac{\partial L_\text{inner}}{\partial t}=-\tfrac{1}{4}N_\text{sol} \bar{V}_\text{SEI}=\frac{c_\text{sol,0}D_\text{sol}(T)\bar{V}_\text{SEI}}{4L_\text{outer}},\\
    \frac{\partial L_\text{outer}}{\partial t}=-\tfrac{1}{4}N_\text{sol} \bar{V}_\text{SEI}=\frac{c_\text{sol,0}D_\text{sol}(T)\bar{V}_\text{SEI}}{4L_\text{outer}}.
\end{gather}
In all other equations, $L_\text{SEI}$ can be substituted with $L_\text{inner}+L_\text{outer}$. For example:
\begin{equation}\label{eta_SEI}
    \eta_\text{SEI}=\rho_\text{SEI}(L_\text{inner}+L_\text{outer})\frac{j_\text{tot}}{a_-}.
\end{equation}

\begin{figure}[h]
	\centering
	\includegraphics[width=0.5\linewidth]{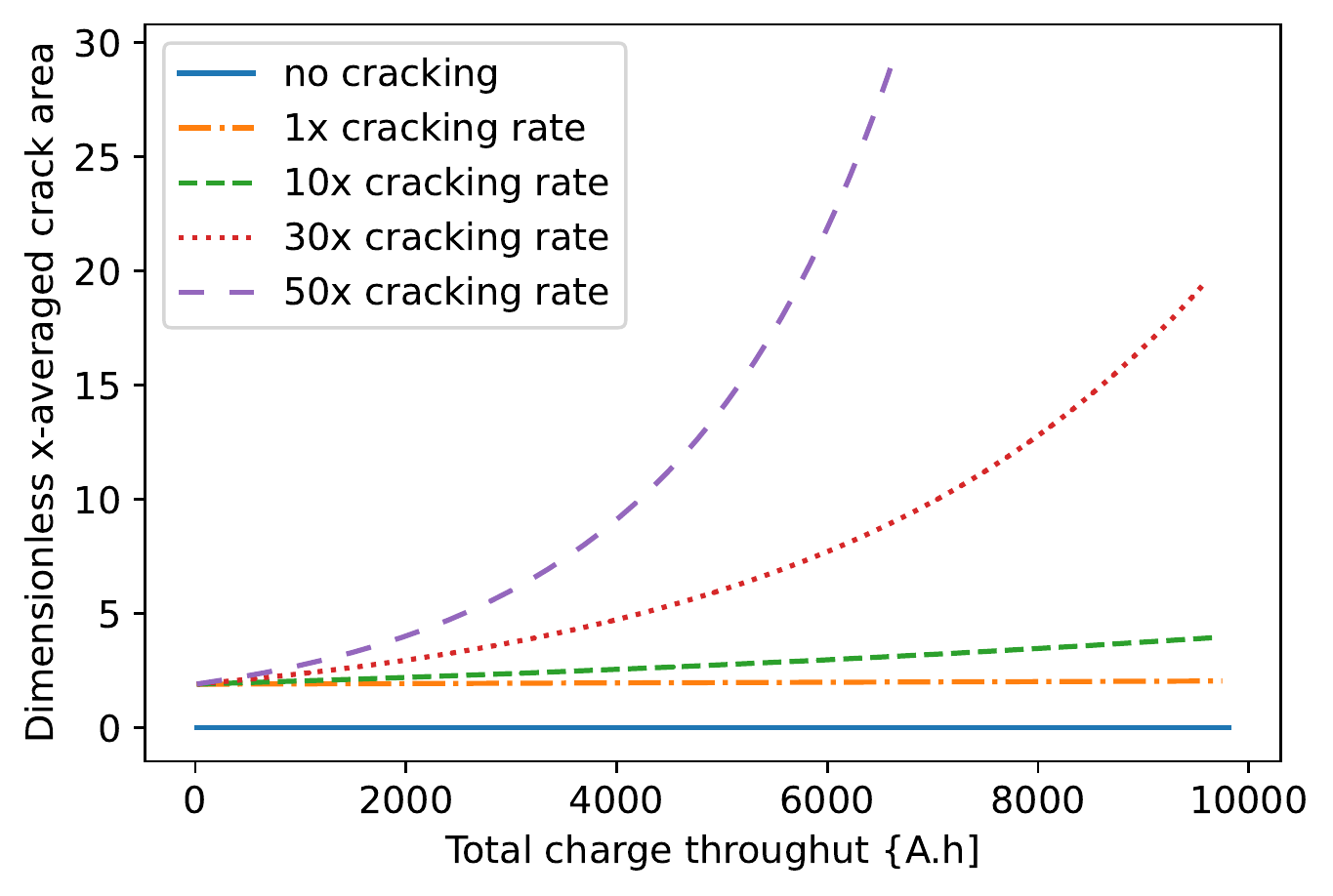}
	\caption{\hl{Increase in dimensionless crack area (interfacial area of cracks} divided by interfacial area without cracks) during the parametric study for particle cracking.}
	\label{roughness} 
\end{figure}
\begin{figure}[h]
	\centering
	\includegraphics[width=0.8\linewidth]{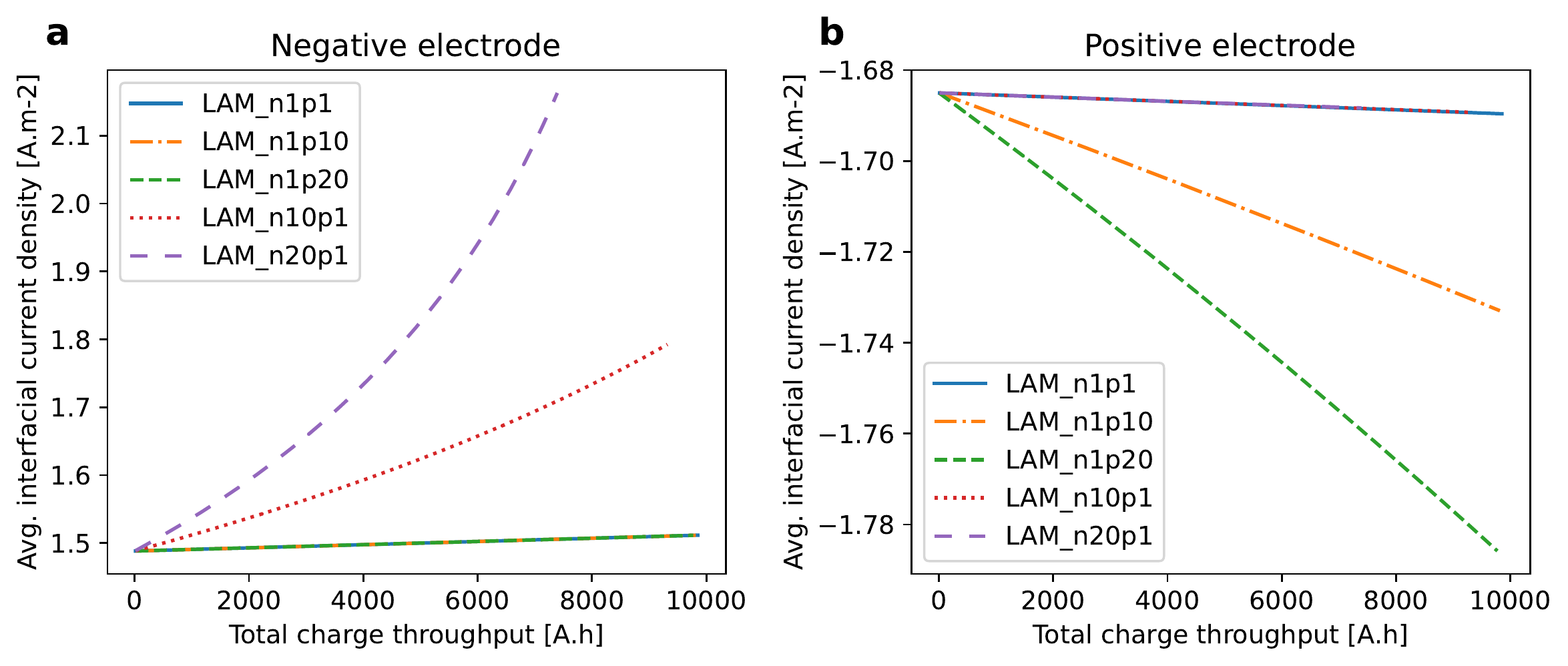}
	\caption{Influence of loss of active material on the magnitude of averaged interfacial current density during battery discharge in the: (a) negative electrode and (b) positive electrode.}
	\label{LAM_interfacial} 
\end{figure}
\begin{figure}[h]
    \centering
    \includegraphics[width=\linewidth]{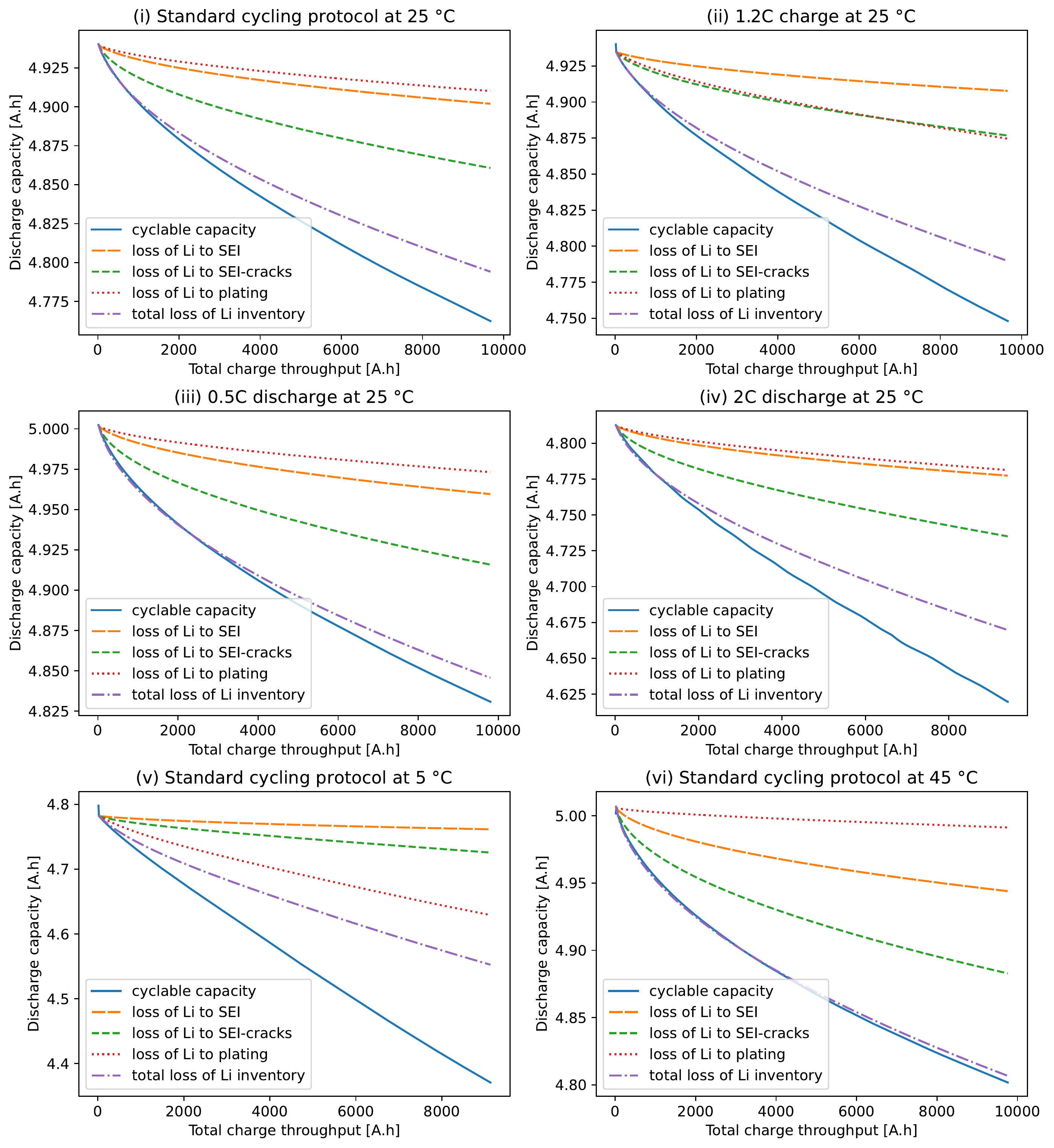}
    \caption{Loss of lithium inventory for cycling protocols (i)-(vi), with contributions from each mechanism: surface SEI, SEI on cracks and lithium plating. The cyclable capacity is also shown for comparison.}
    \label{LLI_plots}
\end{figure}

\bibliographystyle{unsrt}
\bibliography{references}